\documentclass[twocolumn,aps,pra,amsmath,amssymb,superscriptaddress]{revtex4}

\usepackage{bm,color,bbm}
\usepackage{soul}  
\usepackage{hyperref,mathtools,graphicx}

\newcommand{\ie}{{\em i.e.}}

\newcommand{\beq}{\begin{equation}}
\newcommand{\eeq}{\end{equation}}
\newcommand{\bqa}{\begin{eqnarray}}
\newcommand{\eqa}{\end{eqnarray}}
\newcommand{\nn}{\nonumber}

\newcommand{\erf}[1]{Eq.~(\ref{#1})}

\newcommand{\sch}{Schr\"odinger}

\newcommand{\sq}[1]{\left[ {#1} \right]}
\newcommand{\cu}[1]{\left\{ {#1} \right\}}
\newcommand{\ro}[1]{\left( {#1} \right)}
\newcommand{\an}[1]{\left\langle{#1}\right\rangle}

\newcommand{\tp}{^{\top}}

\newcommand{\del}{{\boldsymbol \nabla}} 
\newcommand{\du}{\partial} 

\newcommand{\dbd}[2]{\frac{\partial{#1}}{\partial{#2}}}

\newcommand{\at}[2]{\left.{#1}\right|_{#2}}

\newcommand{\red}{\color{red}}

\newcommand{\blk}{\color{black}}
\newcommand{\blu}{\color{blue}}
\definecolor{maroon}{rgb}{0.7,0,0}

\definecolor{ngreen}{rgb}{0.2,0.6,0.2}
\newcommand{\grn}{\color{ngreen}}
\definecolor{golden}{rgb}{0.8,0.6,0.1}

\renewcommand\grn{\blk}
\renewcommand\blu{\blk}
\renewcommand\red{\blk}   

\begin{document}
\title{Quantum phenomena modelled by interactions between many classical worlds}  
\author{Michael J. W. Hall} 
\affiliation{Centre for Quantum Dynamics, Griffith University, Brisbane, QLD 4111, Australia}
\author{Dirk-Andr\'e Deckert}
\affiliation{Department of Mathematics, University of California Davis, One Shields Ave, CA 95616, USA}
\author{Howard M. Wiseman}\email{H.Wiseman@Griffith.edu.au}
\affiliation{Centre for Quantum Dynamics, Griffith University, Brisbane, QLD 4111, Australia}

\begin{abstract}
    We investigate whether quantum theory can be understood as the continuum
    limit of a mechanical  theory, in which there is a huge, but finite, number
    of classical `worlds', and quantum effects arise solely from a universal
    interaction between these \blk 
    worlds, without reference to any wave function.  Here a `world' means an
    entire universe with well-defined properties, determined by the classical
    configuration of its  particles and fields.  
    In  our approach each world evolves deterministically; probabilities arise
    due to ignorance as to which world a given observer occupies;  and we argue
    that in the limit of infinitely many worlds the wave function can be
    recovered (as a secondary object) \blk from the motion of these worlds.  We 
    introduce a simple model of such \blk a `many interacting worlds' approach 
    and show \blk that it can reproduce some
    generic quantum phenomena---such as Ehrenfest's theorem, wavepacket
    spreading, barrier tunneling and zero point energy---as a direct consequence
    of mutual repulsion between worlds. 
    Finally,  we perform numerical simulations using our approach. We demonstrate, first,
    that it can be used to calculate \blk 
    quantum ground states, \blk and second,
    that it is capable of  reproducing, at least qualitatively, the double-slit \blk interference phenomenon. \blk
\end{abstract}

\maketitle

\section{Introduction}

The role of the wave function differs markedly in various formulations of
quantum mechanics.  For example, in the Copenhagen  interpretation it is a
necessary tool for calculating statistical correlations between \blk {\em a
priori} \blk classical preparation and registration devices \cite{copenhagen};
in the de~Broglie-Bohm interpretation it acts as a pilot wave that guides the
\blk world's classical \blk configuration \cite{dbb}; in the
many-worlds interpretation it describes an ever-branching tree of noninteracting
\blk quasiclassical \blk worlds \cite{mwi}; and in spontaneous collapse models
its objective `collapse' \blk creates a single quasiclassical world \blk 
\cite{grw}.  

In other formulations the wave function does not even play a primary role.  For
example, in Madelung's quantum hydrodynamics \cite{madelung}, Nelson's
stochastic dynamics \cite{nelson}, and Hall and Reginatto's exact uncertainty
approach \cite{exact}, the fundamental equations of motion are formulated in
terms of a configuration probability density $P$ and a momentum potential $S$
(or the gradient of the latter), with a purely formal translation to a
wave function description via $\Psi:=P^{1/2}\exp[iS/\hbar]$. \blk These
approaches can describe the evolution of any scalar wave function on
configuration space, which includes any fixed number of spinless particles, plus
bosonic fields.   In this paper we will similarly
treat spinless and bosonic degrees of freedom. \blk 

More recently, it has  been observed by Holland \cite{holland} and by Poirier and
coworkers \cite{poirier, parlant,schiff} that \blk the evolution of  such \blk
quantum systems 
can be formulated without reference
even to a momentum potential $S$.  Instead, nonlinear Euler-Lagrange equations
are used to define trajectories of a continuum of fluid elements, in an
essentially hydrodynamical picture.  The trajectories are labelled by a
continuous parameter, such as the initial position of each element, and the
equations involve partial derivatives of up to fourth order with respect to this
parameter. 

In the Holland-Poirier \blk hydrodynamical \blk approach the wave function plays
no dynamical role.  However, it may be recovered, in a nontrivial manner, by
integrating the trajectories up to any given time \cite{holland}.  This has
proved a useful tool for making efficient and accurate numerical calculations in
quantum chemistry \cite{poirier,parlant}.  \blk Schiff and Poirier
\cite{schiff}, while ``drawing no definite conclusions'', interpret their
formulation as a ``kind of ``many worlds'' theory'', albeit they have a
continuum of trajectories (\ie~flow lines), not a discrete set of worlds. \blk 

Here we take a different \blk but related \blk approach, with the aim of
avoiding the ontological difficulty of a continuum of worlds.  In particular, we
explore the possibility of replacing the continuum of fluid-elements
in the Holland-Poirier approach by a \blk huge \blk but
finite number of interacting \blk `worlds'.  Each world is classical in the
sense  \blu of having \blk determinate properties \blu that \blk are functions of its
configuration.
In the absence of the interaction with
other worlds, each world evolves according to classical Newtonian physics.  All
quantum effects arise from, and only from, the interaction between worlds. We
therefore call this the `many interacting worlds' (MIW) approach to quantum
mechanics.  A broadly similar idea has been independently suggested by Sebens
\cite{sebens}, although without any explicit model being given.

The MIW approach \blk can only become \blk equivalent to standard quantum
dynamics in the continuum limit, where the number of worlds becomes \blk
uncountably \blk infinite.  However, we will show that even  in  the case of just two
interacting worlds \blk it \blk is a useful toy model for modeling and
`explaining' quantum phenomena, such as the spreading of  wavepackets  and
tunneling through a potential barrier. \blk
Regarded as a fundamental physical theory
in its own right, the MIW approach may also lead to new predictions arising from
the restriction to a finite number of worlds. \red Finally, \blk it provides a natural
discretisation of the Holland-Poirier approach which may be useful for numerical
purposes. \red Before considering how its dynamics might be mathematically 
formulated and used as a numerical tool, however, we give a brief discussion of how 
its ontology may appeal to those who favour realist interpretations. \blk

\subsection{Comparative ontology of the MIW approach} 

\grn At the current stage, the MIW approach is \red not yet well enough 
developed to be considered on equal grounds with other long-established 
realistic approaches to quantum mechanics such as  the de~Broglie-Bohm (dBB) and
many-worlds (MW) interpretations. Nevertheless we think it is  of interest \red to compare its ontology with those of these better known approaches. 
\blk 

\blk In the MIW approach \red there is no wave-function, only a very large number 
of classical-like worlds with definite configurations \blk that
evolves deterministically. Probabilities arise only because observers are
ignorant of which world they actually occupy, and so assign an equal weighting
\blk to all worlds compatible with the  macroscopic state of affairs they
perceive. In a typical quantum experiment, where the outcome is indeterminate in
orthodox quantum mechanics, the final configurations of the worlds in the MIW
approach can be grouped into different classes based on macroscopic properties
corresponding to the different possible outcomes. The orthodox quantum
probabilities will then be approximately proportional to the number of worlds in
each class. \blk 

In contrast, the dBB interpretation postulates a single \red classical-like \blk world, deterministically
guided by a physical \red universal \blk wave function.  This world \red --- a single
point of configuration space --- \blk does not exert any back reaction on the
guiding wave, which has no source but which occupies the entire configuration
space. This  makes it \red challenging \blk to give an ontology for the wave function in
parts of configuration space so remote from the `real' configuration that it will
never affect \red its trajectory \blk (a nice analogy can be found in Feynman's criticism of classical
electromagnetism \cite{feynman}). 
Furthermore, \blk this wave function also
determines a probability density for the initial world configuration
\cite{dbb,hollandbook}.  From a Bayesian perspective this dual role is not easy to reconcile
\cite{wiseman07}.
 
In the \blk Everett or \blk MW interpretation, the `worlds' are orthogonal
components of a universal wave function \cite{mwi}. \blk The particular
decomposition at any time, and the identity of worlds through time is argued to
be defined (at least well-enough for practical purposes) by the quantum dynamics
which generates essentially independent evolution of these quasiclassical worlds
into the future (a phenomenon called effective decoherence). The inherent
fuzziness of Everettian worlds is in contrast to the
corresponding concepts in the MIW approach, of a  well-defined group \blk of
deterministically-evolving configurations. \blk In the MW interpretation \blk it
is meaningless to ask exactly how many worlds there are at a given time, or
exactly when a branching event into subcomponents occurs, leading to criticisms
that there is no precise ontology \cite{kent}.  Another \blk difficult issue
 is that worlds are not \blk equally `real' in the MW
interpretation, but are `weighted' by the \blk modulus squared of the \blk
corresponding superposition coefficients. \blk As noted above,  in the MIW
approach all worlds are equally weighted, so that Laplace's theory of
probability is sufficient to account for our experience and expectations. \blk  
 
 \red The skeptical reader may wonder whether it is appropriate to call 
 the entities in our MIW theory ``worlds'' at all. After all, each ``world'' corresponds 
 to a set of positions of particles in real (3D) space. How, then, is the nature and 
 interaction of these worlds any different from those of different gas species, say 
 $A$ and $B$, where the positions of all the $A$ molecules
 constitute one ``world'' and those of the $B$ molecules (each one partnered, 
 nominally and uniquely, with one of the $A$ molecules) 
constitute another ``world''? The answer lies in the details of the interaction.
 
 In the above example, any given $A$ molecule 
will interact with any $B$ molecule whenever they are close together in 
3D space. Thus a hypothetical being in the ``$A$ world'', 
made of some subset of the $A$ molecules, 
would experience the presence of $B$ molecules in much the same way that it would
feel the presence of other $A$ molecules.  By contrast, as will be shown later in this 
paper, the force between worlds in our MIW approach is non-negligible only when the 
two worlds are close in {\em configuration} space. It would be as if the $A$ gas 
and $B$ gas were completely oblivious to each other unless every single $A$ molecule 
were close to its $B$ partner. Such an interaction is quite unlike anything in classical 
physics, and it is clear that our hypothetical $A$-composed observer would have no 
experience of the $B$ world in its everyday observations, but by careful experiment 
might detect a subtle and nonlocal action on the $A$ molecules of its world. 
Such action, though involving very many, rather than just two, worlds, is what we propose could 
lie behind the subtle and nonlocal character of quantum mechanics. \blk

\subsection{Outline of the paper}

In Sec.~\ref{sec:MIW} we introduce the MIW approach in detail.  
The approach allows flexibility in specifying the precise dynamics of the interacting worlds. Hence,
we consider 
a whole class of MIW models which, as we discuss in Sec.~\ref{sec:limit},
should \blk agree with the predictions of orthodox \blk quantum mechanics in the
limit of infinitely many worlds.  We present a particularly simple model \blk in
Sec.~\ref{sec:example} and demonstrate in Sec.~\ref{sec:phenomena} that it is
capable of reproducing some \blk well-known quantum phenomena: Ehrenfest's \blk
theorem, wavepacket spreading, barrier tunneling, and zero-point energy. We
conclude with a numerical analysis of oscillator ground-states and double-slit
interference phenomena in Secs.~\ref{sec:groundstates} \blk and
\ref{sec:quantum-evolution}, respectively.\blk

\section{Formulation of the many interacting worlds approach}

\subsection{From dBB to MIW}\label{sec:MIW}

For pedagogical reasons we introduce MIW from the perspective of the dBB interpretation of quantum mechanics, and hence, start with a brief review of the latter.  
We regard the MIW approach as fundamental, but to show that it should (in an appropriate limit) 
reproduce the predictions of orthodox quantum mechanics, 
it is convenient to build up to it via the equations of dBB mechanics.
It is well known that dBB interpretation reproduces all 
the predictions of orthodox quantum mechanics, so far as the latter is well defined \cite{hollandbook, BerZan05}.

Consider then a universe comprising $J$  scalar nonrelativistic distinguishable
particles, in a \blk $D$-dimensional space. (As mentioned in the introduction we
could also include bosonic fields, but for simplicity we omit them here.) In dBB
mechanics there is a universal (or `world') wave function $\Psi_t({\bf q})$
defined on configuration space. Here the argument of wavefunction is a vector
${\bf q} = \cu{ {q}^1 , \cdots ,{q}^{K} }\tp$ of length $K=D{J}$.  We order
these variables so that  for $D=3$, the vector $(q^{3j-2},q^{3j-1},q^{3j})\tp$
describes the position of the $j$th \blk particle. For convenience we associate
a mass with each direction in configuration space, so that the mass of the $j$th
particle appears here 3 times, as $m^{3j-2} = m^{3j-1} = m^{3j}$. \blk Then we
can write \sch's  equation for the world wave function $\Psi_t({\bf q})$ as 
\beq 
\label{schr}
i\hbar \dbd{}{t}\Psi_{t}({\bf q}) = \sq{ \sum_{k=1}^{K} \frac{\hbar^2}{2 m^k} \ro{\dbd{}{q^k}}^2 + 
V({\bf q}) } \Psi_t({\bf q}).
\eeq 
As well as the world wave function, there is another part of the dBB ontology,
corresponding to the real positions of all the particles. We call this the {\em
world-particle}, with position ${\bf x}(t) = \cu{ {x}^1(t) , \cdots ,{x}^{K}(t)
}\tp$, which we also call the world configuration.  
Reflecting  ``our ignorance of the  initial conditions''~\cite{dbb}, \blk 
the initial position ${\bf x}(0)$ of the world-particle is a random
variable distributed according to probability density \blk $P_0({\bf x})$, \blk
where 
\beq
P_t({\bf q}) = |\Psi_{t}({\bf q})|^2.
\eeq
The velocity of the world-particle, $\dot {\bf x}$, \blk is then defined by  
\beq \label{velocityS}
m^k \dot{x}^k(t) =  \at{\dbd{S_t({\bf q})}{q^k}}{{\bf q}={\bf x}(t)}
\eeq
where 
\beq \label{sdef}
S_t({\bf q}) = \hbar \arg[\Psi_t({\bf q})].
\eeq 
This equation of motion guarantees that the probability   density \blk for the world-configuration 
${\bf x}(t)$ at any time $t$ is given by $P_t({\bf x})$. This property is known 
as {\em equivariance} \cite{hollandbook,quantumequilibrium}.

In Bohm's original formulation \cite{dbb}, the law of motion (\ref{velocityS})  
 is expressed equivalently by the second-order equation 
 \beq
m^k\ddot{x}^k = f^k({\bf x}) + r^k_t({\bf x}),\label{Bohmacc}
\eeq
with \erf{velocityS} applied \blu as \blk a constraint on the velocity at the initial time ($t=0$). \blk 
Here the force has been split into classical (${\bf f}$) and quantum (${\bf r}$)   contributions, 
the latter called ${\bf r}$ because of its locally repulsive nature, which will be shown \blk later. \blk 
These are \blk
 defined by 
\beq
{\bf f}({\bf q})  = - {\del V({\bf q})}, \ \  
{\bf r}_t({\bf q}) =  -{\del  Q_t({\bf q})}   \label{Bohmforce}
\eeq
(with the $k$th component of $\del$ being $\du/{\du q^k}$).
Here 
\beq Q_t({\bf q}) =  [P_{t}({\bf q})]^{-1/2} \sum_{k=1}^K \frac{-\hbar^2}{2m^k} \ro{\dbd{}{q^k}}^2 [P_{t}({\bf q})]^{1/2}, 
\label{eq:quantum-potential} 
\eeq
was called the quantum potential by Bohm \cite{dbb}, \blk and vanishes for $\hbar = 0$. 

It can be shown that \erf{Bohmacc} reproduces \erf{velocityS} at all times. 
Although \erf{Bohmforce} looks like Newtonian mechanics there is no conserved energy for the world-particle  alone, because $Q_t$ is time-dependent in general. Moreover, the wave function $\Psi_{t}({\bf q})$ evolves in complete indifference to the world particle, so there is no transfer of energy there. Thus, these dynamics are quite unlike those familiar from classical mechanics. \blk

Suppose now that instead of only one \blk world-particle, as in the dBB interpretation, 
there  were a huge number ${N}$ of world-particles co-existing, with positions (world-configurations)  
${\bf x}_1, \ldots, {\bf x}_n, \ldots, {\bf x}_N$. If each of the $N$   initial \blk
world-configurations is chosen at random   from $P_0({\bf q})$ \blk
\blk
as described above, then 
\beq \label{p0q}
P_0({\bf q}) \approx N^{-1}\sum_{n=1}^{N}\delta\ro{{\bf q} -{\bf x}_n(0)} 
\eeq
by construction.  The approximation is in the statistical sense that the averages of any sufficiently smooth function ${\varphi}({\bf q})$, calculated via either side, will be approximately equal, and it becomes arbitrarily good in the limit $N\to\infty$.  Clearly, a similar approximation also holds if the empirical density on the right hand side is replaced by a suitably smoothed version thereof.
\blk
By equivariance the quality of   of either \blk approximation is then conserved 
for all times. 

One can thus approximate \blk  $P_t({\bf q})$, and its derivatives, from 
a suitably smoothed version of the empirical density at time $t$. From this   smoothed density \blk one   may \blk also 
obtain a corresponding approximation of the Bohmian force (\ref{Bohmforce}) 
\begin{align}
{\bf r}_t({\bf q})  \approx {\bf r}_N({\bf q};{\bf X}_t) \blk \qquad \text{for }N\gg 1 ,
  \label{eq:quantum-force}
\end{align}
  in terms of the list of world-configurations \blk
\beq
{\bf X}_t=\cu{{\bf x}_1(t), {\bf x}_2(t), \ldots {\bf x}_N(t) } 
\eeq
at time $t$.  Note in fact that since only local properties of $P_t({\bf q})$ are required 
for ${\bf r}_t({\bf q})$, the approximation ${\bf r}_N({\bf q};{\bf X}_t)$ requires 
only worlds from the set ${\bf X}_t$ which are in the $K$-dimensional {\em neighbourhood} 
of ${\bf q}$. That is, the approximate force is local in $K$-dimensional
configuration space. 

\blk We now take the crucial step of
replacing the Bohmian force \blk (\ref{Bohmforce}), which acts on each
world-particle ${\bf x}_n(t)$ via (\ref{Bohmacc}), by the approximation ${\bf
r}_N({\bf x}_n(t);{\bf X}_t)$. Thus, the evolution of world-configuration ${\bf
x}_n(t)$ is directly determined by the other configurations in ${\bf X}_t$.
This makes the wave function $\Psi_t({\bf q})$, and the functions $P_t({\bf
q})$ and $S_t({\bf q})$ derived from it, superfluous. \blk
What is left is a mechanical theory, referred to as MIW, which describes the
motion of a `multiverse' of ${N}$ co-existing worlds ${\bf x}_1(t), \ldots {\bf
x}_n(t), \ldots {\bf x}_N(t)$, where each world-configuration ${\bf x}_n(t)$ is
a $K$-vector specifying the position of $\blk J\blk=K/D$ particles.

\blk While the MIW approach has been motivated above as an approximation to the
dBB interpretation of quantum mechanics, we have the opposite in mind. We would
like to regard MIW as the fundamental theory, from which under certain
conditions dBB can be recovered as an effective theory provided $N$ is sufficiently
large; see Sec.~\ref{sec:limit} below. \blk
Note that \blk the MIW
approach is conceptually and mathematically very different from dBB.  Its
fundamental dynamics \blk are described by the system of $N\times J \times D$
second-order differential equations 
\beq \label{eq:MIW}
m^k\ddot{x}_n^k(t) = f^k({\bf x}_n(t)) + r^k_N({\bf x}_n(t);{\bf X}_t).
\eeq
In
the absence of an interworld interaction, corresponding to the classical limit
$r^k_N({\bf x}_n(t);{\bf X}_t)=0$ in  Eq.~(\ref{eq:MIW}), the worlds evolve
independently under purely Newtonian dynamics. Hence, all quantumlike effects
arise from the existence of this interaction. It will be seen in Secs.~III-VI
that this nonclassical interaction corresponds to a repulsive force between
worlds having close \blk configurations, leading to a simple and intuitive
picture for many typical quantum phenomena.

Note also that we use the term `MIW approach' rather than `MIW interpretation.'
This is because while some predictions of quantum mechanics, such as the
Ehrenfest theorem and rate of  wavepacket  spreading, will be seen to hold
precisely for any number of worlds, other predictions can be accurately
recovered only  under certain conditions \blk in the limit $N\to\infty$.  This has two immediate implications:
the possibility of experimental predictions different from standard quantum
mechanics, due to the finiteness of $N$, and the possiblity of using the MIW
approach for approximating the dynamics of standard quantum systems in a
controlled manner. In the latter case  Eq.~(\ref{eq:MIW}) \blk must be
supplemented by suitable initial conditions, corresponding to choosing the
initial world-configurations randomly from $P_0$, and the initial world-particle
velocities from $S_0$ via  Eq.~(\ref{velocityS}) \blk (see also below and
 Sec.~\ref{sec:quantum-evolution}\blk).
\blk

\subsection{Probabilities and the quantum limit}\label{sec:limit}

While each world evolves deterministically under  Eq.~(\ref{eq:MIW})\blk, which
of the ${  N\blk}$ worlds we are actually living in, compatible with the
perceived macroscopic state of affairs, is unknown.  Hence, assertions about
the configuration of the  $J$ \blk particles in our world naturally become
probabilistic, as it was the case in dBB.  In particular, for a given function
${\varphi}({\bf x})$ of the world configuration, only an equally-weighted {\em
population} mean \blk  
\begin{align} \label{expec}
&\an{\varphi({\bf x})}_{\bf X} \equiv \blk \frac{1}{N} \sum_{n=1}^N {\varphi}({\bf x}_n) ,
\end{align}
over all the worlds compatible with observed macroscopic properties, can be
predicted at any time.

We shall now show that under certain conditions the MIW expectation
values (\ref{expec}) are expected to converge to the ones predicted by quantum theory 
when the number of worlds $N$ tends to infinity.
For this let us suppose we are provided a solution to the
\sch\ equation $\Psi_t$ on $K$-dimensional configuration space, 
and the initial configurations of the
$N$ worlds, ${\bf x}_1(0),{\bf x}_2(0),\dots,{\bf x}_N(0)$ are approximately
distributed according to the distribution $P_0({\bf q})=|\Psi_0({\bf
q})|^2$.  
Hence, at $t=0$ one has, for any smooth function $\varphi$ on configuration space, \blk 
\begin{align*} 
\langle {\varphi}\blk \rangle_{\Psi_0} &\equiv \int d{\bf q}\,|\Psi_0({\bf q})|^2 \, {\varphi}({\bf q}) \blk \\
 &\approx \frac{1}{N}\sum_{n=1}^N {\varphi}({\bf x}_n(0)) = \an{\varphi({\bf x})}_{{\bf X}_{0}}, \blk
\end{align*}
with the approximation becoming arbitrarily good as
$N\to\infty$ for $\varphi$ sufficiently regular. \blk 
Suppose further that $\Psi_t$ is such that  at initial time $t=0$ the
velocities of the worlds fulfill   Eq.~(\ref{velocityS})\blk.
In consequence of (\ref{eq:quantum-force}),  \blk the trajectory of
the $n$th world generated by (\ref{eq:MIW}) should stay close to the
corresponding Bohmian trajectory generated by   Eq.~(\ref{Bohmacc})\blk. Hence, by
equivariance of the latter trajectories, the world configurations ${\bf
x}_1(t),\dots,{\bf x}_N(t)$ at time $t$ will be approximately distributed
according to the distribution $P_t({\bf x})=|\Psi_t({\bf x})|^2$. That is, \blk
\begin{align} \nn
\langle {\varphi}({\bf x})\rangle_{\Psi_t} &= \int d{\bf q}\,|\Psi_t({\bf q})|^2 {\varphi}({\bf q}) \blk \\ 
\label{timeav}
&\approx \frac{1}{N}\sum_{n=1}^N {\varphi}({\bf x}_n(t)) = \an{\varphi({\bf x})}_{{\bf X}_t} 
\end{align} \blk
as desired. 

In summary, the configuration-space \blk expectation values of the MIW approach should \blk coincide with 
those for the quantum state $\Psi_t$ as
$N\to\infty$ provided that \blk at some time 
$\tau\blk$ the following conditions are
met:
\begin{enumerate}
    \item The worlds ${\bf x}_1(\tau\blk),\ldots,{\bf x}_N(\tau\blk)$ are
        $|\Psi_{\tau\blk}|^2$-distributed;
    \item The velocities of the worlds fulfill Eq.~(\ref{velocityS}) at $t=\tau\blk$.
\end{enumerate}
Since configuration space expectation values are all that is required to establish empirical 
equivalence with orthodox quantum mechanics (just as in dBB~\cite{dbb}), this 
establishes the viability of the MIW approach.  
		
It is a remarkable feature of the MIW approach that only a simple equal
weighting of worlds, reflecting ignorance of which world an observer occupies,
appears to be sufficient to reconstruct quantum statistics in \blu a \blk suitable
limit.  Similarly, in this limit, the observer should see statistics as
predicted by quantum mechanics when carrying out a sequence of experiments in
his or her {\it single} world.  In particular, from the typicality
    analysis by D\"urr {\it et al.} for the dBB interpretation
    \cite{quantumequilibrium}, it follows that Born's rule holds for dBB
    trajectories belonging \blu to \blk typical initial configurations, where in dBB
    typical stands for almost all with respect to the $|\Psi_t|^2$ measure.  Hence,
    since the MIW trajectories are expected to converge to dBB trajectories and
    the world configurations are $|\Psi_t|^2$ distributed in the limit
    described above, Born's rule will hold for typical MIW worlds, where we
    emphasise again that in our MIW approach typicality simply means for the
    great majority of worlds, since each world is equally weighted.

\blu
\subsection{Which initial data gives rise to quantum behaviour?}

 We \grn have argued \blu above \grn that, given a solution to the Schr\"odinger equation,  one
can generate \grn corresponding initial data  for \grn the MIW equations of motion
\eqref{eq:MIW} whose solution approximates quantum theory as $N\to\infty$. Suitable \grn initial data   is as per \grn
conditions 1.~and 2.~above.
\red This suggests a converse question: \grn given a solution, \red ${\bf X}(t)=\cu{{\bf x_1}, \ldots , {\bf x}_N}$ 
\grn to the MIW equations of motion
\eqref{eq:MIW}, \red is \grn there is any solution, $\Psi_t$, to the Schr\"odinger
equation that can be approximately  generated \red by ${\bf X}(t)$? 

The sense in which an  \blu approximation $\tilde{\Psi}_t \red \approx \Psi_t$ \red could be  generated \red is the following. 
\grn Given the world configurations, ${\bf X}(t)$, one can \blu construct approximations  $|\tilde{\Psi}_t|^2$ for the probability density 
 as per \grn Eq.~\eqref{p0q}.    Further, from \grn the velocities, \red $\dot {\bf X}(t)$, \blk 
 approximations $\arg \tilde \Phi_t$ for the phase can be constructed via  \grn  Eqs.~\eqref{velocityS} \blu and \grn \eqref{sdef}.   Thus, \red 
there are many ways a quantum
mechanical candidate for $\tilde{\Psi}_t$ could \grn be constructed.  The relevant measure of
the quality of the approximation at time $t$ is then given by the $L_2$ norm
\begin{align}
    \label{eq:MIW_QM_approx} \|\tilde{\Psi}_t-U(t-\tau)\tilde{\Psi}_\tau\|_2,
\end{align}
where $U(t)$ denotes the corresponding Schr\"odinger evolution and $\tau$ is
some initial time. \blk

\red From the above discussion, it might seem obvious that to obtain \blu approximate \red quantum evolution 
(in this sense) one  must \blu simply \red impose the 
velocity constraint \eqref{velocityS} \blu and \red \eqref{sdef} at \blu the intial time \red $t=\tau$, for some $\Psi_\tau$. 
However, on further reflection, one realises that this is no constraint at all. For any 
finite number $N$ of worlds, there is always some complex function $\Phi({\bf q})$ 
such that setting $\Psi_\tau({\bf q}) = \Phi({\bf q})$ will match the initial 
velocities at the positions of those worlds. The point is that this $\Phi$ may fail to yield 
an approximate solution at later times, so that $\|\tilde{\Psi}_t-U(t-\tau){\Phi}\|_2$ 
is not small, for any approximate reconstruction $\tilde{\Psi}_t$ and $t-\tau$ sufficiently 
large. Thus, a more relevant constraint may be that the velocity constraint 
\eqref{velocityS} \blu and \red \eqref{sdef} holds for some $\Psi_\tau$ that is smoothly varying on 
the scale of the maximum inter-world distance in configuration space.  
We emphasize, however, that it is not completely obvious that such a constraint is necessary \blu---i.e., quantum behaviour may be typical in the MIW approach as $N\rightarrow\infty$---\red 
a point to which we return in the concluding section \blu (see also Sec.~IV). \blk

\subsection{Interworld interaction potential}
\label{sec:class}

The general MIW approach described in Sec.~II~A is only complete once the form of the force ${\bf r}_N({\bf x};{\bf X})$ between worlds, in Eq.~(\ref{eq:MIW}), is specified. There are different possible ways of doing so, each leading to a different version of the approach. \red However, \blk it is natural to seek a formulation of the MIW approach in which the interaction force between worlds is guaranteed to be conservative. That is, a force \blk of the form 
\begin{equation} \label{eq:consforc}
{\bf r}_N({\bf x}_n;{\bf X}) =  -\nabla_{{\bf x}_n} U_N({\bf X})
\end{equation}
for some potential function $U_N({\bf X})$ defined on the $N$ world configurations ${\bf X}=({\bf x}_1,\dots,{\bf x}_N)$, where $\nabla_{{\bf x}_n}$ denotes the gradient vector with respect to ${\bf x}_n$.  

\red If we have an  interworld interaction potential $ U_N({\bf X})$, \blk this immediately allows the equations of motion (\ref{eq:MIW}) to be rewritten in the equivalent Hamiltonian form
\begin{equation} \label{hameq}
\dot{{\bf x}}_n = \nabla_{{\bf p}_n} H_N({\bf X,P}),~~~\dot{{\bf p}}_n = - \nabla_{{\bf x}_n} H_N({\bf X,P}),
\end{equation}
with all its attendant advantages. Here ${\bf P}=({\bf p}_1,\dots,{\bf p}_n)$ defines the momenta of the worlds, 
 with components 
 \beq
 p^k_n=m^k\dot x^k_n,
 \eeq \blk and the Hamiltonian is given by
\begin{equation} \label{ham}
H_N({\bf X,P}) := \sum_{n=1}^N\sum_{k=1}^K \frac{(p^k_n)^2}{2m^k} + \sum_{n=1}^N V({\bf x}_n) + U_N({\bf X}).
\end{equation}
We will refer to $ U_N({\bf X})$ as the interworld interaction potential.

To motivate the form of suitable interworld potentials $U_N$, note that, to reproduce quantum mechanics in the limit $N\to\infty$, \blu it is natural to require that \blk the average energy per world, $\an{E}_{N} \equiv N^{-1}H_N({\bf X,P})$,  \blk
approaches the quantum average energy in this limit.  
 Note that we cannot associate a definite energy with each world, because of the interworld interaction. 
 Note also that we have substituted the subscript $N$ for the subscript ${\bf X}$ 
 (or ${\bf X},{\bf P}$ as would be needed in this case), for ease of notation.\blk  
 
 Now, if the configurations ${\bf x}_1,\dots,{\bf x}_N$ sample the distribution $|\Psi({\bf x})|^2$, the quantum average energy can be written, using $\Psi=P^{1/2}\exp[iS/\hbar]$, as \blk \cite{dbb}
\begin{align} \label{avq} \nn
\langle E\rangle_{\Psi} &= \int d{\bf q}\, P({\bf q})\left[\sum_{k=1}^K \frac{1}{2m^k}\left(\frac{\partial S}{\partial q^k}\right)^2+ V({\bf q}) \right.\\ \nn
&~~~~~~~~~~~~~~~~~~~~~~~~+ \left. \sum_{k=1}^K \frac{\hbar^2}{8m^k P^2} \left( \frac{\partial P}{\partial q^k}\right)^2 \right]  \\ \nn
&\approx \frac{1}{N}\sum_{n=1}^N  \left[\sum_{k=1}^K \frac{1}{2m^k}\left(\frac{\partial S}{\partial q^k}\right)^2+ V({\bf q}) \right.\\ \nn
&~~~~~~~~~~~~~~~~~~~~~~~~+ \left. \left. \sum_{k=1}^K \frac{\hbar^2}{8m^k P^2} \left( \frac{\partial P}{\partial q^k}\right)^2 \right]  \right|_{{\bf q}={\bf x}_n}
\end{align}
(for $N$ sufficiently large).  Moreover, the average energy per world is given
via Eq.~(\ref{ham}) \blk as 
\begin{align} \nn
\an{E}_{N} \blk &= N^{-1}H_N({\bf X},{\bf P}) \\
&= \frac{1}{N}\sum_{n=1}^N\left[\sum_{k=1}^K \frac{(p^k_n)^2}{2m^k} +  V({\bf x}_n)\right] + \frac{1}{N} U_N({\bf X}). \nn
\end{align}
Assuming that the trajectories generated by  the Hamiltonian $H_N$  are close to the dBB trajectories for sufficiently large $N$, then  $p^k_n=m^k\dot{x}^k_n\approx \partial S({\bf x}_n)/\partial x^k_n$, and comparing the two averages shows that $U_N({\bf X})$ should be chosen to be of the form
\begin{equation} \label{un}
U_N({\bf X}) = \sum_{n=1}^N \sum_{k=1}^K \frac{1}{2m^k}   \left[ g_N^k({\bf x}_n;{\bf X})\right]^2 ,
\end{equation}
where 
\begin{equation} \label{gn} 
g_N^k({\bf q} ;{\bf X})  \approx \frac{\hbar}{2}\frac{1}{P({\bf q})} \frac{\partial P({\bf q})}{\partial q^k} .
\end{equation}
Here the left-hand side is to be understood as an approximation of the right-hand side, \blk 
obtained  via a suitable smoothing of the empirical density in  
Eq.~(\ref{p0q})\blk , analogous to the approximation of the quantum force $r_N({\bf q})$ by ${\bf r}_N({\bf x};{\bf X})$ in Eq.~(\ref{eq:quantum-force}). 
It is important to note that a good approximation of the \blu quantum \blk force (which is essential to obtain QM in the large $N$ limit), \blu via Eqs.~(\ref{eq:consforc}) and~\ref{un}), \blk
is not guaranteed by a good approximation in \erf{gn}. \blu This is easy to see  in the case that
the empirical density is smoothed using only members of a (large) fixed subset of the $N$ worlds, implying that the  \blu
force in~(\ref{eq:consforc}) will vanish for all worlds not contained in this subset. \red For example, 
in the $K=1$ case considered below, one could choose the subset of  odd-numbered 
worlds (\ie~$n$ odd), in which case the even-numbered worlds would behave classically.  
\blk Thus \erf{gn} is a guide only, and \erf{eq:quantum-force} must also be checked. \blk

The interworld potential $U_N$ in Eq.~(\ref{un}) is positive by construction.
This  leads directly to the existence of a minimum energy and a corresponding
stationary configuration for the $N$ worlds, corresponding to the quantum
groundstate energy and (real) groundstate wavefunction $P^{-1/2}({\bf q})$ \blk
in the limit $N\to\infty$.  Note that $g_N^k({\bf q} ;{\bf X})$ in
Eq.~(\ref{gn}) approximates Nelson's osmotic momentum \cite{nelson},  suggesting that $U_N$ in Eq.~(\ref{un})
may be regarded as a sum of nonclassical kinetic energies, as is  
explored in Sec.~IV. The quantity $g_N^k({\bf q} ;{\bf X})$  also approximates the imaginary part of the `complex momentum' that appears in complexified Bohmian mechanics \cite{john,wyatt}. 

Further \blk connections between $U_N({\bf X})$, ${\bf g}_N({\bf q}
;{\bf X})$ and ${\bf r}_N({\bf x};{\bf X})$ are discussed in
Appendix~\ref{apx:A}.

\section{ Simple  example}\label{sec:example}
 
A full specification of the MIW dynamics requires generation of a suitable force
function ${\bf r}_N({\bf x}_n; {\bf X})$  in Eq.~(\ref{eq:MIW})  or a suitable
function $g_N^k({\bf q} ;{\bf X})$ in Eq.~(\ref{un}).  As discussed in the
previous section, these may be viewed as corresponding to approximations of
$-\nabla Q$ and $\nabla P/P$, respectively.
One may obtain many candidates for interworld forces and potentials in this way,
and it is a matter of future interest to determine what may be the most natural
ones.

\grn In this section we demonstrate in the simplest case of one dimension and one
particle per world ($K=D=J=1$) how the proposed MIW approach can be
mathematically substantiated, using a Hamiltonian formulation. \blk
 In
Sec.~IV we show that this example provides a nice toy model for successfully
describing various quantum phenomena.

\subsection{One-dimensional toy model}

For $N$ one-dimensional worlds, it is convenient to distinguish them from the general case by denoting the configuration ${\bf x}_n$ and momentum ${\bf p}_n$ of each world by $x_n$ and $p_n$, respectively.  It is also convenient to label the configurations such that $x_1< x_2<\dots <x_N$ (it will be seen that this ordering is preserved by the repulsive nature of the interaction between worlds).  Similarly, we will use $X = (x_1, \ldots, x_N)$ in place of ${\bf X}$, 
and $P$ in place of ${\bf P}$. Note that we can thus consider $X$ and $P$ as vectors. \blu The mass of each one-dimensional world particle will be denoted by $m$. \blk 

The aim here is to obtain a simple form for the interworld potential $U_N$ in Eq.~(\ref{un}).  It is easiest to first approximate the empirical distribution of worlds by a smooth probability density, $P_{X}(q)$, and use this to obtain a suitable form for $g_N^k({q} ;{X})$ in Eq.~(\ref{gn}).

Now, any smooth interpolation $P_{X}(q)$, of the empirical density of the $N$ worlds, must satisfy,
\begin{align*} 
\frac{1}{N} \sum_{n=1}^N { \varphi}(x_n) &\approx \int dq\,P_{X}(q)\,{ \varphi}(q)\\
&\approx \sum_{n=1}^N \int_{x_{n-1}}^{x_{n}} dq\, P_{X}(x_n)\,{ \varphi}(x_n)\\
& = \sum_{n=1}^N (x_{n}-x_{n-1})\,P_{X}(x_n)\,{ \varphi}(x_n)
\end{align*}
 for sufficiently slowly varying functions ${ \varphi}(x)$ 
 and sufficiently large $N$.
 This suggests the following ansatz  
\begin{equation} \label{papprox}
 P_{X}(x_n) = \frac{1}{N(x_{n}-x_{n-1})} \approx \frac{1}{N(x_{n+1}-x_n)} 
\end{equation}
for the smoothed distribution $P_X$ \blu at $x=x_n$, \blk  which relies on an assumption that the 
inter-world separation is slowly varying. The success of this ansatz requires that the dynamics 
we derive based upon it preserve this slow variation, which appears to be the case from 
the simulations in Sec.~\ref{sec:quantum-evolution}.  
Under this assumption, the two expressions in \erf{papprox} have a relative difference $O(1/N)$. 
Further, for large $N$, we can approximate the derivative of $P_{X}(q)$ at $q=x_n$ by
\begin{equation} \label{toygrad} 
P'_{X}(x_n)\approx \frac{P_{X}(x_{n+1}) - P_{X}(x_n)}{x_{n+1}-x_n}. 
\end{equation}
Hence, $\nabla P_{X}/P_{X}$ is approximated by
\begin{align*} 
\frac{P'_{X}(x_n)}{P_{X}(x_n)} &\approx N[P_{X}(x_{n+1}) - P_{X}(x_n)]\\
&\approx \frac{1}{x_{n+1}-x_n} - \frac{1}{x_{n}-x_{n-1}} .
\end{align*}

It follows that one may take the interworld potential $U_N({X})$ in Eq.~(\ref{un}) to have the rather simple form, 
\begin{equation} \label{un1d}
U_N({X}) = \frac{\hbar^2}{8m} \sum_{n=1}^N \left[ \frac{1}{x_{n+1}-x_n} - \frac{1}{x_{n}-x_{n-1}} \right]^2 .
\end{equation}
To ensure the summation is well defined for $n=1$ and $n=N$, we formally define $x_0=-\infty$ and $x_{N+1}=\infty$. 

\subsection{Basic properties of the toy model}\label{sec:prop}

The MIW toy model corresponding to Eq.~(\ref{un1d}) is defined by the equations
of motion (\ref{hameq}) and Hamiltonian (\ref{ham}).  Since the total energy is
conserved, two adjacent worlds $x_n$ and $x_{n+1}$ cannot approach each other
arbitrarily closely, as the potential $U_N$ would become unbounded. This implies
a mutually repulsive force that acts between neighbouring worlds, \blk
which will be seen to be responsible for a number of generic quantum effects in
Sec.~IV.   Note also that $U_N$ is invariant under translation of the
configurations.  This leads directly to an analog of the quantum Ehrenfest
theorem, also shown in Sec.~IV. 

While of a simple form, the interworld potential $U_N$ in Eq.~(\ref{un1d}) is seen to be a sum of three-body terms, rather than the more usual two-body interactions found in physics.    Further, the corresponding force,  acting on configuration $x_n$, may be evaluated as 
\begin{align}
{r}_N({x_n};{X}) &= -\partial U_N({X})/\partial x_n \nn \\
& = \frac{\hbar^2}{4m} [\sigma_{n+1}({X}) - \sigma_n({X})],  \label{hamforce}
\end{align}
\red which actually involves five worlds, since $\sigma_n({X})$ \blk involves \blu four \blk worlds:  
\begin{align}
\sigma_n({X}) &= \frac{1}{(x_{n}-x_{n-1})^2}\left[ \frac{1}{x_{n+1}-x_{n}} - \frac{2}{x_{n}-x_{n-1}} \right. \nn \\ 
&~~~~~~~~~~~~~~~~~~~~~~~~~~~~~\left. 
+\, \frac{1}{x_{n-1}-x_{n-2}}\right] , \label{sigmaforce}
\end{align}\blk 
(defining $x_n=-\infty$ for $n<1$ and $=\infty$ for $n>N$).

\blk It is shown in
Appendix~\ref{apx:A} that this force corresponds to a particular approximation of the
Bohmian force $r_t(x_n)$, as per Eq.~(\ref{eq:quantum-force}).

\section{Quantum phenomena as generic MIW effects}\label{sec:phenomena}

Here we show the MIW approach is capable of reproducing some  well-known
quantum phenomena, including Ehrenfest's theorem, wavepacket spreading, barrier
tunneling, zero-point energy, and a Heisenberg-like uncertainty relation. We primarily use the toy model of Sec.~\ref{sec:example} for this purpose, although a number of the results, such as the Ehrenfest theorem and wavepacket spreading, are shown to hold for broad classes of MIW potentials. 

\subsection{Ehrenfest theorem: translation invariance}

For any MIW potential $U_N$ in Eq.~(\ref{ham}) that is invariant under translations of the world configurations, such as the toy model in Eq.~(\ref{un1d}), one has
\begin{equation} \label{ehrenfest}
\frac{d}{dt}\langle {\bf x}\rangle_N = \frac{1}{m}\langle {\bf p} \rangle_N,~~~~\frac{d}{dt}\langle {\bf p}\rangle_N = -\langle \nabla V\rangle_N,
\end{equation}
 where $\langle {\varphi}({\bf x},{\bf p}) \rangle_N:= N^{-1}\sum_{n=1}^N
 {\varphi}({\bf x}_n,{\bf p}_n)$.  Thus, the quantum Ehrenfest theorem 
 corresponds to \blk the special case $N\rightarrow\infty$, with the more
 general result holding for any value of $N$.  For example, when $V({\bf x})$ is
 no more than quadratic  with respect to the components of ${\bf x}$, the
 centroid of any phase space trajectory follows the classical equations of
 motion, irrespective of the number of worlds.

To prove the first part of Eq.~(\ref{ehrenfest}), note  from Eqs.~(\ref{hameq}) and (\ref{ham}) that
\begin{align*}
\frac{d}{dt}\langle {\bf x}\rangle_N &= \frac{1}{N} \sum_{n=1}^N \dot{{\bf x}}^n = \blu \frac{1}{N} \sum_{n=1}^N \nabla_{{\bf p}_n} H_N \blk \\
&= \frac{1}{N} \sum_{n=1}^N \frac{1}{m} {\bf p}_n = \frac{1}{m}\langle {\bf p} \rangle_N
\end{align*}
as required.  To prove the second part, note one also has from Eqs.~(\ref{hameq}) and (\ref{ham}) that
\[ \frac{d}{dt}\langle {\bf p}\rangle_N = -\langle \nabla V\rangle_N -\frac{1}{N}\sum_{n=1}^N \nabla_{{\bf x}_n} U_N({\bf X}). \]
Now, translation invariance is the condition 
\beq 
U_N({\bf x}_1+{\bf y},\dots,{\bf
x}_N+{\bf y})=U_N({\bf x}_1,\dots,{\bf x}_N), 
\eeq
\blk and taking the gradient thereof
with respect to ${\bf y}$ at ${\bf y}=0$  yields  $\sum_{n=1}^N
\nabla_{{\bf x}_n} U_N({\bf X})=0$.  The second part of Eq.~(\ref{ehrenfest})
immediately follows.

\subsection{Wavepacket spreading: inverse-square scaling}

It is well known that the position variance of a free one-dimensional quantum particle of mass $m$, with initial wave function $\Psi_0$, increases quadratically in time \cite{merz}:
\[ {\rm Var_{\Psi_t}} x = {\rm Var_{\Psi_0}}x + \frac{2t}{m}{\rm Cov}_{\Psi_0}(x,p) + \frac{2t^2}{m}\!\left[\langle E\rangle_{\Psi_0}- \frac{\langle p\rangle_{\Psi_0} ^2}{2m}\right]\!. \]
Here  $\langle E\rangle_{\Psi}$ and  $\langle p\rangle_{\Psi} $ denote the average energy and momentum, respectively, and ${\rm Cov}_\Psi(x,p)$ denotes the position and momentum covariance, $\langle\Psi| (xp+px)/2|\Psi\rangle - \langle x\rangle_{\Psi} \langle p\rangle_{\Psi}$.  This is often referred to as the spreading of the wavepacket \cite{merz}.

Here we show that an equivalent result holds for $N$ one-dimensional worlds, for any interworld potential $U_N$ satisfying translation invariance and the inverse-square scaling property
\begin{equation} \label{scaling}
U_N(\lambda x_1,\dots,\lambda x_n) = \lambda^{-2}\, U_N(x_1,\dots,x_N) , 
\end{equation}
such as the toy model in Eq.~(\ref{un1d}).  

In particular,  the position variance for $N$ one-dimensional worlds at time $t$ may be written via Eq.~(\ref{expec}) as
\[ {\cal V}_N(t) = \frac{1}{N}\sum_{n=1}^N x_n^2 - \left[\frac{1}{N}\sum_{n=1}^N x_n\right]^2 = \frac{1}{N}{X}\cdot {X} -\langle x\rangle_N^2. \]
Differentiating with respect to time then gives, using Eq.~(\ref{hameq}) and (\ref{ham}) with $V(x)\equiv 0$ and noting that $(d/dt)\langle p\rangle_N=0$ from the Ehrenfest theorem (\ref{ehrenfest}),
\[ \dot{\cal V}_N = \frac{2}{m}\left[ N^{-1} {X}\cdot{P}  - \langle x\rangle_N\,\langle p\rangle_N\right] = \frac{2}{m}{\rm Cov}_{N,t}(x,p), \] 
and 
\[
\ddot{\cal V}_N = \frac{2}{mN}\left[ \frac{P\cdot P}{m} -\sum_{n=1}^N x_n\frac{\partial U_N}{\partial x_n}\right] - \frac{2}{m^2}\langle p\rangle_N^2.
\]
Now, differentiating scaling condition (\ref{scaling}) with respect to $\lambda$, and setting $\lambda=1$, gives $\sum_{n=1}^N x_n (\partial U_N/\partial x_n) = -2U_N$.  Hence, recalling that the average energy per world is $\langle E\rangle_N=N^{-1}H_N$, one obtains
\[ \ddot{\cal V}_N = \frac{4}{m}\langle E\rangle_N - \frac{2}{m^2}\langle p\rangle_N^2 = {\rm constant}. \]
Finally, integration yields the variance at time $t$ to be
\begin{equation} \label{spreadn}
{\cal V}_N(t) = {\cal V}_N(0) + \frac{2t}{m}{\rm Cov}_{N,0}(x,p) + \frac{2t^2}{m}\!\left[\langle E\rangle_N- \frac{\langle p\rangle_N^2}{2m}\right],
\end{equation}
which is of precisely the same form as the quantum case above. Like the generalised Ehrenfest theorem (\ref{ehrenfest}), this result holds for any number of worlds $N$.

The spreading of variance {\it per se} is due to the repulsive interaction
between worlds having close \blk configurations (see previous
section). The above result demonstrates that  a spreading which is \blk
quadratic in time is a simple consequence of an inverse-square scaling property
for the interworld potential.   Such a scaling is generic, since the quantum
potential in Eq.~(\ref{eq:quantum-potential}), that is replaced by $U_N$ in the
MIW approach, also has this property.  Note that it is straightforward to
generalise this result to configuration spaces of arbitrary dimension $K$, with
${\cal V}_N$ replaced by the $K\times K$ tensor $\langle {\bf X}\cdot{\bf X}^T \rangle_N -\langle {\bf
X}\rangle_N \cdot \langle{\bf X}^T\rangle_N$. Here the transpose refers to the 
configuration space index $k$, while the dot-product refers to the world-index $n$ 
as previously.\blk 

\subsection{Barrier tunneling: mutual repulsion}

In quantum mechanics, a wavepacket incident on a potential barrier will be
partially reflected and partially transmitted, irrespective of the height of the
barrier. The probabilities of reflection and transmission are dependent on the
energy of the wavepacket relative to the energy of the barrier \cite{merz}.  In
the MIW formulation, the same qualitative behaviour arises as a generic
consequence of mutual repulsion between worlds having close \blk
configurations.  This is investigated here analytically for the toy model in
Sec.~III, for the simplest possible case of just two worlds, $N=2$. 

\subsubsection{Nonclassical transmission}

\begin{figure}[!t]
	\centering
		\includegraphics[width=0.47\textwidth]{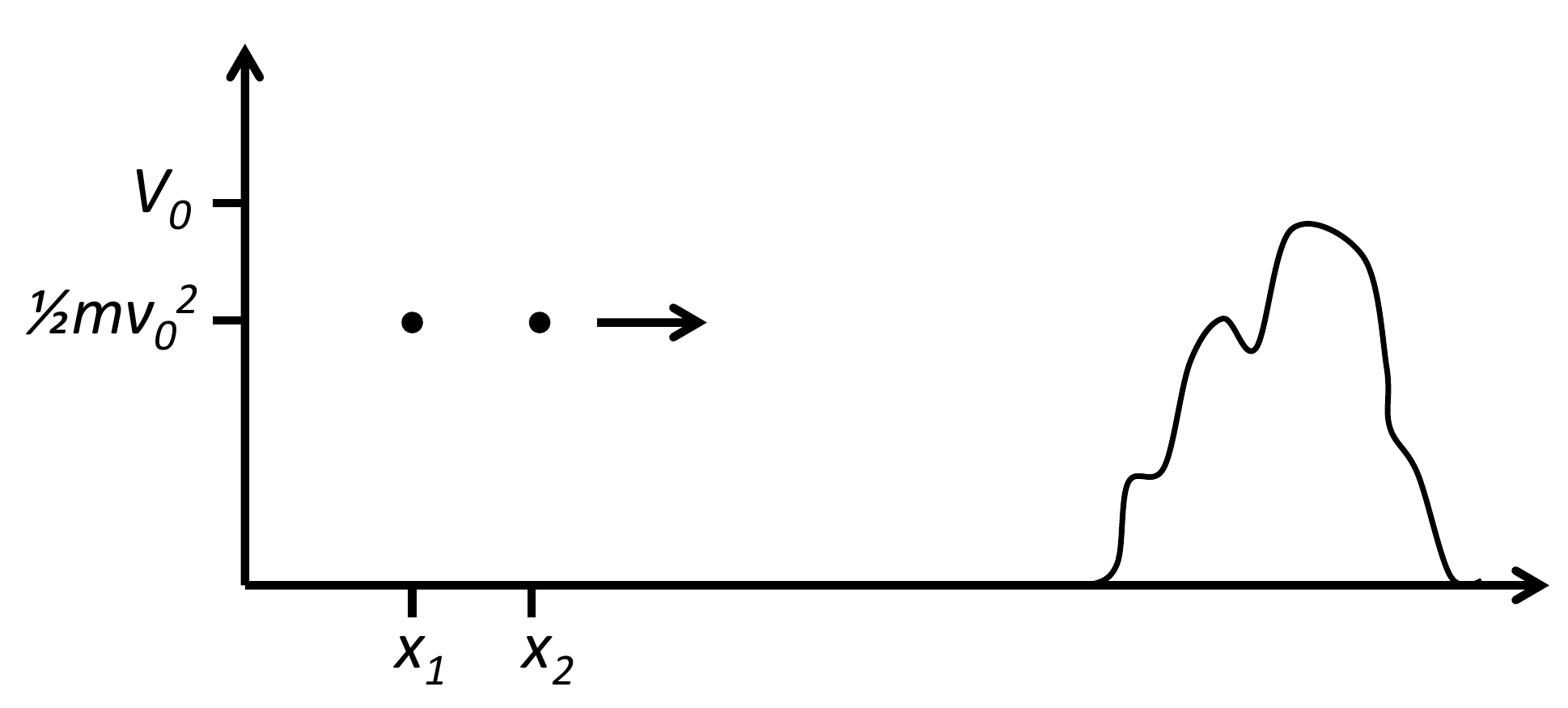}
	
\caption{Two one-dimensional worlds, with equal masses $m$ and initial speeds $v_0$, approach a potential energy barrier of height $V_0$. Due to mutual repulsion the leading world can gain sufficient kinetic energy to pass the barrier. \label{fig:barrier}}
\end{figure}

Consider two one-dimensional worlds, the configurations of which initially approach a potential barrier $V(x)$ from the same side, with kinetic energies  less than the height of the barrier (Fig.~\ref{fig:barrier}).  In the absence of an interaction between the worlds, the configurations will undergo purely classical motion, and so will be unable to penetrate the barrier.  However, in the MIW  approach the mutual repulsion between worlds will boost the speed of the leading world.  This boost can be sufficient for this world to pass through the barrier region, with the other world being reflected, in direct analogy to quantum tunneling.

To demonstrate this analytically, consider the  toy model  in Sec.~III for $N=2$.  The  Hamiltonian simplifies via Eqs.~(\ref{ham}) and (\ref{un1d}) to
\begin{align} \label{toyham}
H = \frac{p_1^2+p_2^2}{2m} + V(x_1) + V(x_2) +\frac{\hbar^2}{4m(x_1-x_2)^2},
\end{align}
where $V(x)=0$ outside the barrier and has a maximum value $V_0>0$ in the barrier region. For simplicity we will restrict to the case of equal initial velocities $v_0$ in the direction of the barrier, with $\frac{1}{2}mv_0^2<V_0$ (Fig.~\ref{fig:barrier}).  Thus, in the classical limit $\hbar  = 0$, penetration of the barrier is impossible.

Defining relative and centre-of-mass coordinates by $q:=x_2-x_1$, $\tilde{q}:=(x_2+x_1)/2$, with conjugate momenta $p=(p_2-p_1)/2$ and $\tilde{p}=p_1+p_2$, respectively, then while the configurations of each world are outside the barrier region their evolution is governed by the Hamiltonian
\[ H_0(q,\tilde{q},p,\tilde{p}) = \frac{\tilde{p}^2}{2M} + \frac{p^2}{2\mu} + \frac{\hbar^2}{8\mu} \frac{1}{q^2}, \]
with initial conditions  $p(0)=0$ and $\tilde{p}(0) = Mv_0$, where $M=2m$ and $\mu=m/2$.  Thus, the equations of motion for $q$ and $\tilde{q}$ decouple, and may be integrated to give
\begin{align} \label{cons}
\tilde{q}(t)=\tilde{q}_0 + v_0 t,~~~~\frac{1}{2}\dot{q}^2 +  \frac{\hbar^2}{8\mu} \frac{1}{q^2} = \frac{\hbar^2}{8\mu} \frac{1}{q_0^2} ,
\end{align}
where the $0$ subscript indicates initial values. \blk These are valid up until one of the configurations reaches the barrier region. The second equation may be further integrated to give the exact solution
\begin{align}
q(t) = \left[q_0^2 + (\hbar t/mq_0)^2\right]^{1/2}.
\end{align}
Note that this increasing mutual separation is in agreement with Eq.~(\ref{spreadn}), with ${\cal V}=q^2/4$, 
and is solely due to the inverse-square interaction between the worlds in Eq.~(\ref{toyham}). 

It follows from Eq.(\ref{cons}) that the maximum possible separation speed is
$\dot{q}_\infty:=\hbar/(mq_0)$,
which can be arbitrarily closely reached if the particles start sufficiently far to the left of the barrier. In this case, the speed of the leading particle is well approximated by 
\[ \dot{x}_2 = \dot{\tilde{q}} +\frac{1}{2}\dot{q} \approx v_0+\hbar/(2mq_0)\] 
at the time it reaches the barrier region.  Hence, transmission through the barrier is always possible providing that the corresponding kinetic energy is greater than $V_0$, \ie, if 
\begin{equation} \label{trans} 
v_0 + \frac{\hbar}{2mq_0} > v_{\rm classical} :=\sqrt{2V_0/m} ,
\end{equation}
\blu where \blk $v_{\rm classical}$ is the initial speed required for a classical particle to penetrate the barrier.  

Eq.~(\ref{trans}) \blk clearly demonstrates that one world configuration can pass through the barrier if the initial separation between the worlds, $q_0$, is sufficiently small. In terms of energy conservation, such a small separation gives rise to a correspondingly large interworld potential energy in Eq.~(\ref{toyham}), which is converted into a  kinetic energy sufficiently large for barrier transmission.  Further, the second configuration will suffer a corresponding loss of kinetic energy, leading to its reflection by the barrier.  

It is seen that even the simplest case of just two interacting worlds  provides a toy model for the phenomenon of quantum tunneling, with an interpretation in terms of the energy of repulsion between close \blk configurations. While this case can be treated analytically, the interactions between $N>2$ worlds are more complex.  It would therefore be of interest to investigate the general case numerically, including tunneling delay times.

\subsubsection{Nonclassical reflection}

The same toy model also captures the nonclassical phenomenon that a barrier can reflect a portion of a quantum wavepacket, even when the incident wavepacket has a large average kinetic energy \cite{merz}.  In particular, consider the case where the initial kinetic energies of both worlds are  larger than the barrier height for the `toy' Hamiltonian in Eq.~(\ref{toyham}), \ie, $mv_0^2/2>V_0$.  Hence, in the classical limit $\hbar = 0$ both configurations will pass through the barrier region. 

It follows from the above analysis that the leading configuration is always transmitted --- the kinetic energy of this world is only further increased by the mutual repulsion energy.  However, choosing the initial distance from the barrier to be large enough for the separation speed to approach $\dot q_\infty$, it follows that the second configuration will not be transmitted  if
\begin{equation} \label{ref} 
v_0 - \frac{\hbar}{2mq_0} < \sqrt{2V_0/m} = v_{\rm classical}.
\end{equation}
Indeed, it will not even reach the barrier region if the left hand side is less than zero. This result demonstrates the converse to Eq.~(\ref{trans}): it is always possible for one configuration to be reflected, if the intial separation between the worlds is sufficiently small, as a consequence of being slowed by the mutual repulsion between worlds.

\subsection{Zero point energy: mutual exclusion}

The classical groundstate energy of a confined system corresponds to zero momentum and a configuration that minimises the classical potential $V({\bf x})$.  However, the corresponding quantum groundstate energy is always greater, with the difference referred to as the quantum zero point energy.  

For example, for a one-dimensional quantum system  the Heisenberg uncertainty relation $(\Delta x)_\Psi (\Delta p)_\Psi \geq \hbar/2$ implies 
\begin{equation} \label{enboundq}
\langle E\rangle_\Psi \geq \langle V\rangle_\Psi + \frac{\hbar^2}{8m \,{\rm Var}_\Psi x} 
\end{equation}
for the average energy of any state $\Psi$ (noting that $\langle
p^2\rangle_\Psi\geq {\rm Var}_\Psi p$).  Hence, for any state with finite
average energy, the system cannot be confined to a single point, as this would
imply ${\rm Var}_\Psi x =0$. In particular, it cannot be confined to the
classical groundstate position $x_{\rm min}$, corresponding to the classical
groundstate energy  $V_{\rm min}=V(x_{\rm min})$, and therefore $\langle
E\rangle_\Psi >V_{\rm min}$. \blk

In the MIW formulation, a zero point energy similarly arises because no two worlds can be confined to the same position -- and to the classical groundstate configuration ${\bf x}_{\rm min}$ in particular. This mutual exclusion of configurations is a consequence of the repulsion between worlds, which forces the interworld potential $U_N$ in Eq.~(\ref{un}) to be strictly positive.  For example, as shown in  Appendix~\ref{apx:D}, \blk the average energy per world for the  toy model in Sec.~III satisfies 
\begin{equation} \label{enboundn}
\langle E\rangle_N \geq \langle V\rangle_N + \left(\frac{N-1}{N}\right)^2\frac{\hbar^2}{8m  {\cal V}_N} .
\end{equation} 
This is clearly very similar to the quantum bound in Eq.~(\ref{enboundq}), with the latter being precisely recovered in the limit  $N\rightarrow\infty$. 

Remarkably, both the quantum and the MIW lower bounds are saturated by the groundstate of a one-dimensional oscillator.  The quantum case is well known \cite{merz}.  For the  toy model the corresponding groundstate energy is
\beq \label{enosc}
\langle E\rangle_{N,{\rm ground}} = \left( 1 - \frac{1}{N}\right) \blk \frac{1}{2} \hbar\omega , 
\eeq
as demonstrated in  Appendix~\ref{apx:groundstates}. \blk Note that this vanishes in the classical limit $\hbar =0$ (or, alternatively, $N=1$), and approaches the quantum groundstate energy $\frac{1}{2}\hbar\omega$ in the limit $N\to\infty$, as expected. 

\subsection{Heisenberg-type uncertainty relation}

It was noted in Sec.~II~C that the interworld potential in Eq.~(\ref{un}) has
the form of a sum of nonclassical kinetic energies, where the corresponding
`nonclassical momentum' of the $n$th world  is given by ${\bf p}_n^{\rm
nc}:={\bf g}_N({\bf x_n} ;{\bf X})$, with the components of ${\bf g}_N$
explicitly defined in Eq.~(\ref{gn}). \blk For the toy model in
Eq.~(\ref{un1d}), this nonclassical momentum has the explicit form
\begin{equation} 
    \label{eq:pnc}
p_n^{\rm nc} = \frac{\hbar}{2}\left[ \frac{1}{x_{n+1}-x_n} - \frac{1}{x_{n}-x_{n-1}} \right] 
\end{equation}
(defining $x_0=-\infty$ and $x_{N+1}=\infty$ as always).
 
It follows immediately that the average of the nonclassical momentum vanishes, \ie,
$$\langle p^{\rm nc} \rangle_N = \frac{1}{N}\sum_{n=1}^N p_n^{\rm nc} = 0 . $$
Hence, the energy bound for $U_N$ in Eq.~(\ref{unbound}) can be rewritten in the rather suggestive form
\begin{equation} \label{heisen}
(\Delta x)_N\, (\Delta p^{\rm nc})_N \geq \ro{1-\frac{1}{N}} \blk \frac{\hbar}{2} ,
\end{equation}
reminiscent of the Heisenberg uncertainty relation for the position and momentum of a quantum system.  Indeed, a nonclassical momentum observable $p^{\rm nc}_\Psi$ may also be defined for quantum systems, which satisfies the uncertainty relation $\Delta_\Psi x \,\Delta_\Psi p^{\rm nc}_\Psi \geq \hbar/2$, and which implies the usual Heisenberg uncertainty relation \cite{eur}. Thus,  Eq.~(\ref{heisen}) has a corresponding quantum analogue in the limit $N\to \infty$. It would be of interest to further investigate the role of the nonclassical momentum in the MIW approach.

\section{Simulating quantum groundstates}\label{sec:groundstates}

The MIW equations of motion (\ref{hameq}) and (\ref{ham}) in the Hamiltonian formulation imply that the configurations of the worlds are stationary if and only if
\begin{equation} \label{stationary}
{\bf p}_n = 0,~~~~\nabla_{{\bf x}_n} \left[ V({\bf x}_n) + U_N({\bf X})\right]  = 0 
\end{equation}
for all $n$.  In particular, the forces acting internally in any world are balanced by the forces due to the configurations of the other worlds.

Unlike quantum systems, the number of stationary states in the MIW approach
(with $N$ finite) \blk is typically finite.  For example, for two
one-dimensional worlds as per Eq.~(\ref{toyham}), with a classical potential
$V(x)$ symmetric about $x=0$, Eq.~(\ref{stationary}) implies that the stationary
configurations are given by  $x_1=-a$ and $x_2=a$, where $a$ is any positive
solution of $V'(a)=\hbar^2/(16ma^3)$.  For the particular case of a harmonic
oscillator potential, $V(x)=(1/2)m\omega^2 x^2$, there is \blk just one
stationary state, corresponding to $a=\frac{1}{2} (\hbar/m\omega)^{1/2}$.  More
generally, the number of stationary state configurations will increase with the
number of worlds $N$.

The MIW formulation suggests a new approach for numerically approximating the
groundstate wave function and corresponding groundstate energy for a given
quantum system.  In particular, the groundstate probability density is
approximately determined by finding the global minimum of $\sum_{n=1}^N V({\bf
x}_n) + U_N({\bf X})$, for suitably large $N$.  Below we will give and test a
`dynamical' algorithm for doing so.  First, however, for the purposes of benchmarking
this algorithm, we calculate the exact groundstate configurations for the
 toy model of Sec.~III, for a harmonic oscillator potential.

\subsection{Oscillator groundstates: exact MIW calculation}

For the one-dimensional toy model defined in Sec.~III, with harmonic oscillator potential $V(x)=(1/2)m\omega^2 x^2$, it is convenient to define the dimensionless configuration coordinates
\begin{equation} \label{qndef}
{\xi}_n := \sqrt{2m\omega/\hbar} \,\,x_n .
\end{equation}
As shown in Appendix~\ref{apx:groundstates}, the unique groundstate configuration for $N$ worlds is then determined by the recurrence relation
\begin{equation}
\label{recur3}
{ \xi}_{n+1} = { \xi}_n - \frac{1}{{ \xi}_1+\dots+{ \xi}_n},
\end{equation}
subject to the constraints
\begin{equation} \label{cond2}
{ \xi}_1+\dots+{ \xi}_N = 0,~~~({ \xi}_1)^2+\dots ({ \xi}_N)^2 = N-1.
\end{equation}
These equations are straightforward to solve for any number of worlds.

\begin{figure}[!t]
	\centering
		\includegraphics[width=0.47\textwidth]{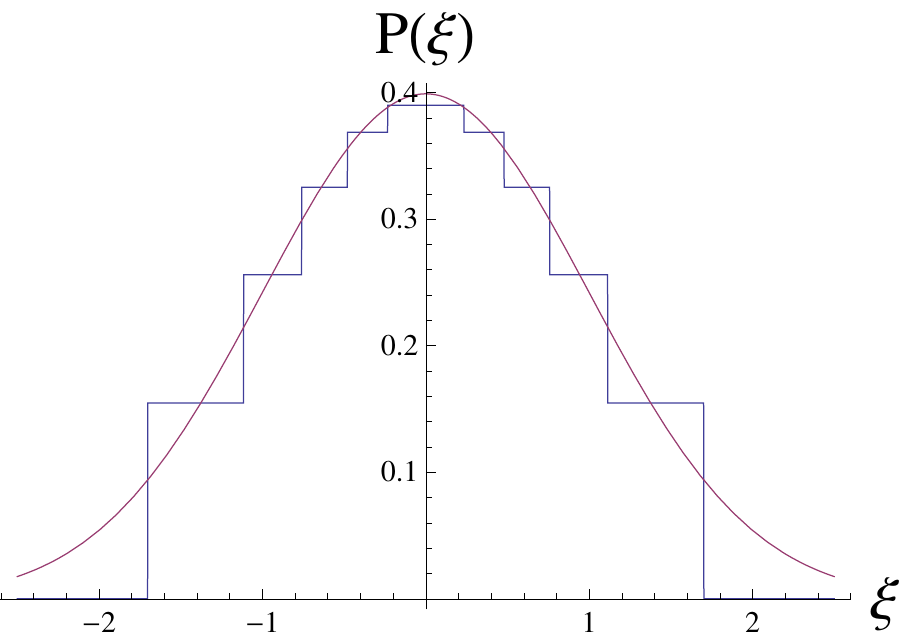}	
\caption{\label{oscground21}(Color online). Oscillator groundstate for $N$=11 worlds.  The steps of
    the stepped blue curve occur at the values $q={ \xi}_1,{
    \xi}_2,\dots,{ \xi}_{11}$, corresponding to the stationary world
    configurations  $x_1,x_2,\dots,x_{11}$ via Eq.~(\ref{qndef}).  The height of
    the step between ${ \xi}_{n}$ and ${ \xi}_{n+1}$ is $P_N({
    \xi}_n):=N^{-1}({ \xi}_{n+1}-{ \xi}_n)^{-1}$, which from
    Eq.~(\ref{papprox}) is expected to approximate the quantum groundstate
    distribution for a one-dimensional oscillator,
    $P_{\Psi_0}(\xi)=(2\pi)^{-1/2}e^{-\xi^2/2}$, for large $N$.  The latter
    distribution is plotted for comparison (smooth magenta curve).
All quantities are dimensionless.}
\end{figure}

As discussed in Appendix~\ref{apx:groundstates}, for small values of 
$N$, the recurrence relation 
can be solved analytically, and in general it can be \blk 
efficiently solved numerically for any given number of worlds $N$.  For example,
for $N=11$ one obtains the configuration depicted in Fig.~\ref{oscground21}. The
corresponding Gaussian probability density for the exact quantum groundstate is
also plotted, as the smooth curve in Fig.~\ref{oscground21}.  It is seen that
the groundstate configuration of just 11 worlds provides a good approximation to
the quantum groundstate for many purposes.  Note also that the corresponding
mean \blk groundstate energy follows from Eq.~(\ref{enosc}) as
  $(5/11)\hbar\omega\approx 0.45\hbar\omega$, which is reasonably close to the
quantum value of $0.5\hbar\omega$.  The approximation improves with increasing
$N$.

\subsection{Oscillator groundstates: dynamical MIW algorithm}

The net force acting on each world is zero for a stationary configuration of
worlds, as per Eq.~(\ref{stationary}). We exploit this fact in the following
algorithm to compute a stationary configuration. \blk

Given an arbitrary configuration of $N$ worlds ${\bf X}(0)$ we iterate the
following two-step algorithm: \blk 1) Set the velocities $\dot{\bf X}(0)$ to zero; 2) 
Integrate Eq.~(\ref{eq:MIW}) over a small time interval $[0,\Delta t]$, and replace the
initial
configuration ${\bf X}(0)$ by
${\bf X}( \Delta t)$. 
Note that each iteration tends to reduce the
total energy from its initial value, as the worlds move away from maxima of the
total potential energy, and the velocities are reset to zero in each
iteration.  The fixed-points of this iterative map are clearly the stationary
states of the MIW equations.   Hence, after a sufficient number of iterations,
the configuration will converge to a stationary configuration.  An alternative algorithm could, instead of setting velocities to zero, add a viscous term to the evolution, in analogy to the dBB-trajectory based algorithm of Maddox and Bittner \cite{mad}.

\begin{figure}[!t]
	\centering
    \includegraphics[width=0.47\textwidth]{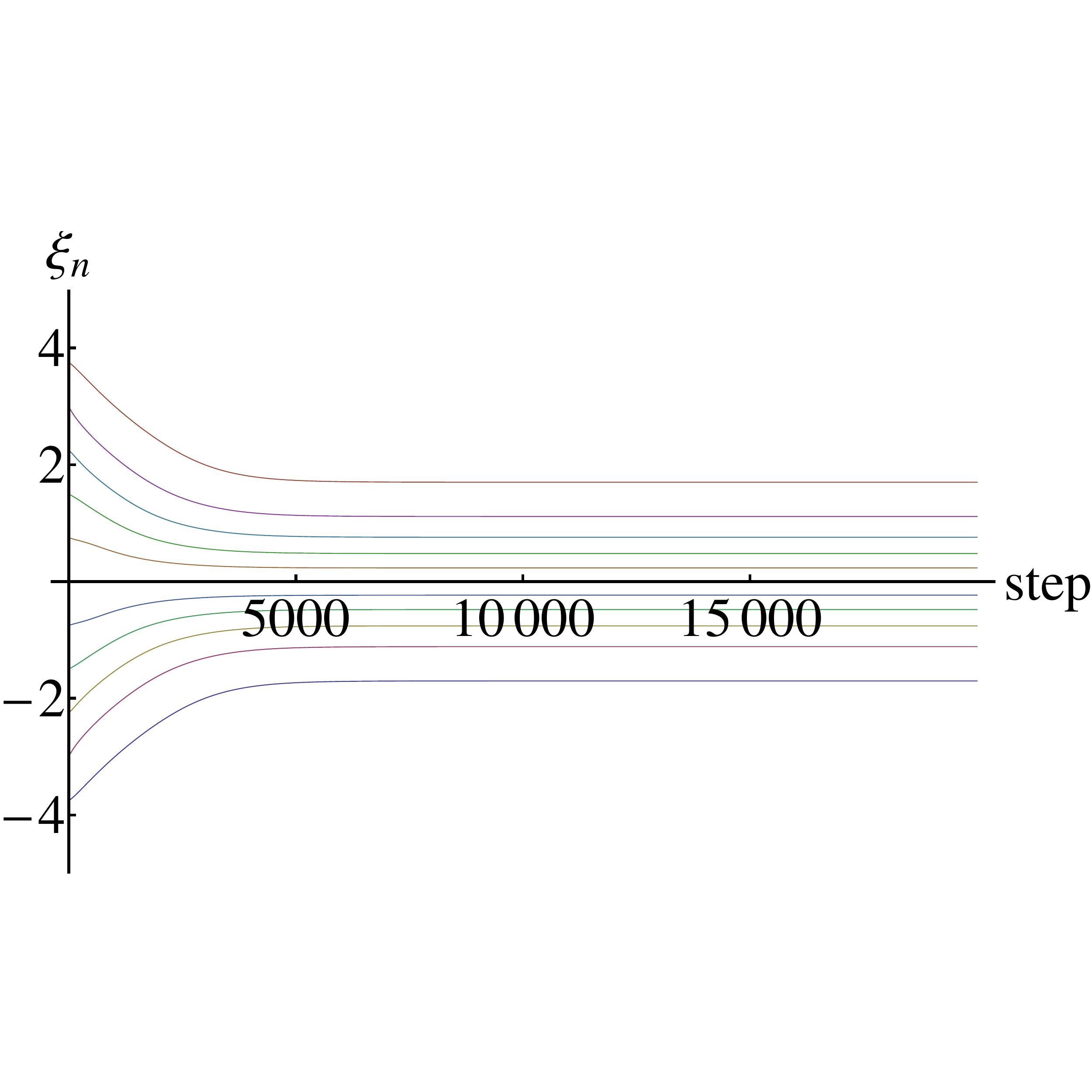}
    \caption{\label{trajectories} (Color online).
  Convergence of the ensemble of 
 world-particles $x_n =(x_1,\ldots,x_N)$ for $N=11$ towards the ground-state configuration
 as a function of the step number in the the iteration algorithm described in the text. 
  Here dimensionless positions are used, $\xi_n = \sqrt{2m\omega/\hbar} \,\,x_n$, as per Eq.~(\ref{qndef}),
  and the temporal step size is 
 $\Delta t = 5\times 10^{-2} \omega^{-1}$.  As the plot shows, convergence is complete by step number 6000, 
  at which the 
        distribution of worlds is close to the Gaussian quantum ground state. }
  \end{figure}
  
    \begin{figure}[!t]
	\includegraphics[width=0.47\textwidth]{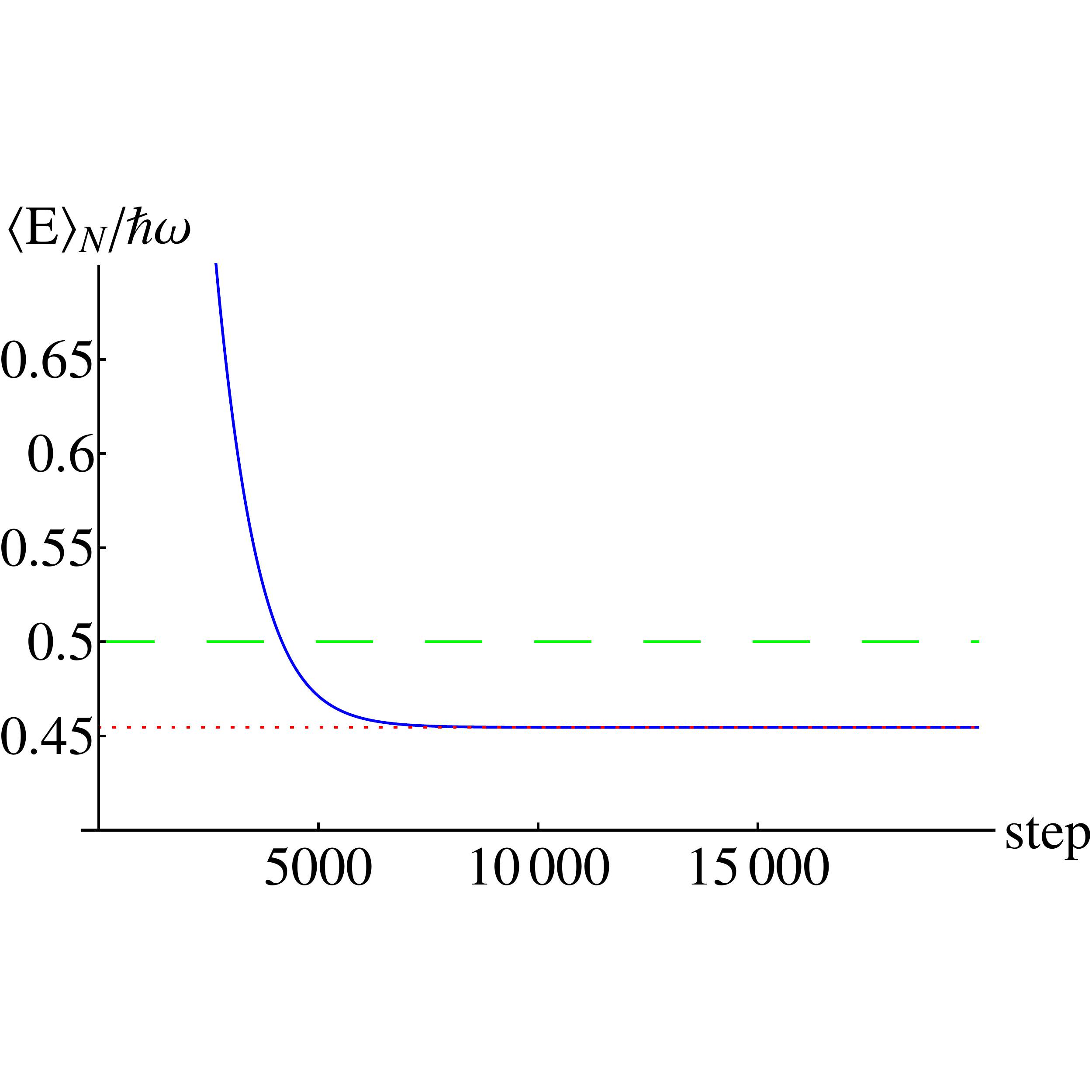}
    \caption{\label{algorithm} (Color online).
 Average energy (solid blue line) for the ensemble of worlds, 
 as a function of iteration step. Details as in Fig.~\ref{trajectories}. 
       Note convergence to the the exact value for MIW is
        $(1-1/N)\hbar\omega/2$ (red \blk dotted line),  which coincides with 
        the quantum mechanical value of $\hbar\omega/2$ (green \blk dashed line) for $N\to \infty$.\blk}
\end{figure}

We have tested this dynamical algorithm \blk for the case of the one-dimensional
harmonic oscillator.  We slightly \blk modified \blk the interworld force in Eq.~(\ref{hamforce}) by
placing auxiliary worlds in fixed positions on either side of the configurations
$x_1<x_x\dots <x_N$ , with two auxiliary worlds to the far left and two to the
far right, rather than at $-\infty$ and $\infty$.  The auxiliary worlds have
only have a tiny effect on the computations, but ensure the interworld force is
well-defined for $n=1,2\dots,N$.   We took advantage of the symmetry of the
oscillator potential $V(x)$ to choose an initial configuration symmetric about
$x=0$.

Convergence of the algorithm was found to be extremely rapid; see 
Figs~\ref{trajectories} and \ref{algorithm}. \blk  For example, on
an laptop running a simple Mathematica implementation of the algorithm, it took
only 30 seconds to converge to the groundstate configuration for the case of
$N=11$ worlds,  with a corresponding groundstate energy accurate to one part in
 $10^{10}$ \blk in comparison to the exact groundstate energy $(5/11)\hbar\omega$
following from Eq.~(\ref{enosc}).

 This \blk dynamical algorithm can be employed to compute ground states 
for all Hamiltonians for which the groundstate wave function has a constant phase. 
\blk The numerical advantage over contemporary techniques to compute groundstate
wave functions is
evident when considering higher dimensional problems, \blk \ie, $K=DJ>1$.
Instead of solving the partial differential equation
$H\Psi_{\rm g}=E_{\rm g}\Psi_{\rm g}$ that lives on a 
$K$-dimensional \blk configuration space (which
memorywise starts to become unfeasible already for three particles in three
dimensions), one only needs to solve
for the stationary state of $K\times \blk N$ coupled ordinary differential equations. The
quantum distribution $|\Psi_{\rm g}|^2$ and energy $E_g$ \blk of the ground state can then be
approximated \blk from the stationary \blk distribution of worlds.

\section{Simulating quantum evolution}\label{sec:quantum-evolution}

Given the apparent success of the MIW approach as a tool for simulating
stationary quantum states (Sec.~V), it is of considerable interest to also
investigate whether this approach can  provide \blk a similarly useful
controlled approximation of the Schr\"odinger   time evolution and, in
particular, whether it is capable of describing quantum interference, which is
one of the most striking quantum phenomena. Here we offer evidence 
that it can, at least for the canonical  `double-slit' problem.

\begin{figure}[!t]
	\centering
	\includegraphics[width=0.47\textwidth]{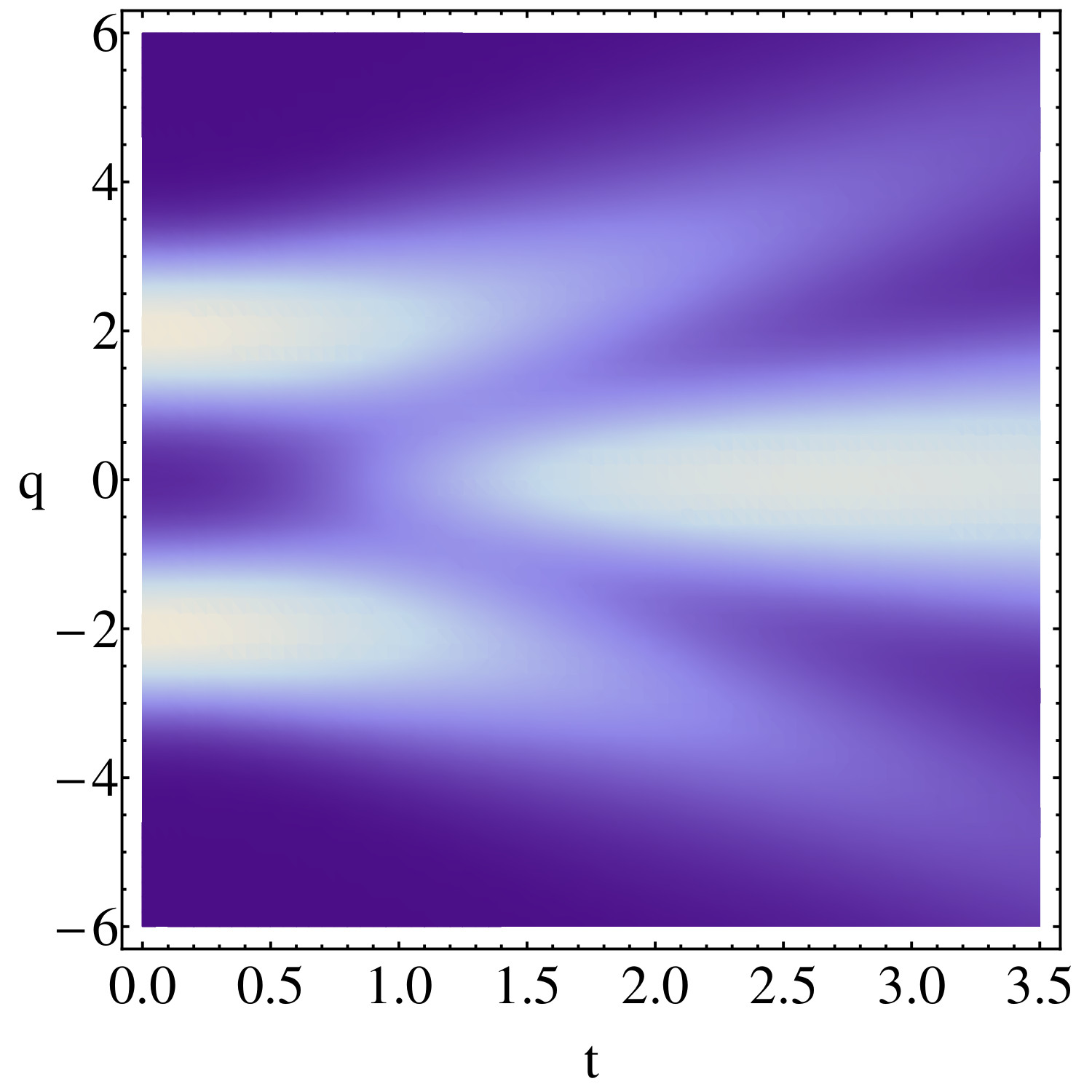}
    \caption{\label{qm} (Color online).
    A density plot of $|\Psi_t|^2$ 
    for the exact quantum evolution of two identical initial Gaussian
    wavepackets in units of the initial spread (\ie~$\sigma=1$); \blk time is given in units
of $\hbar/(2m)$.\blk}
\end{figure}

\begin{figure}[!t]
    \begin{center}
        \includegraphics[width=0.47\textwidth]{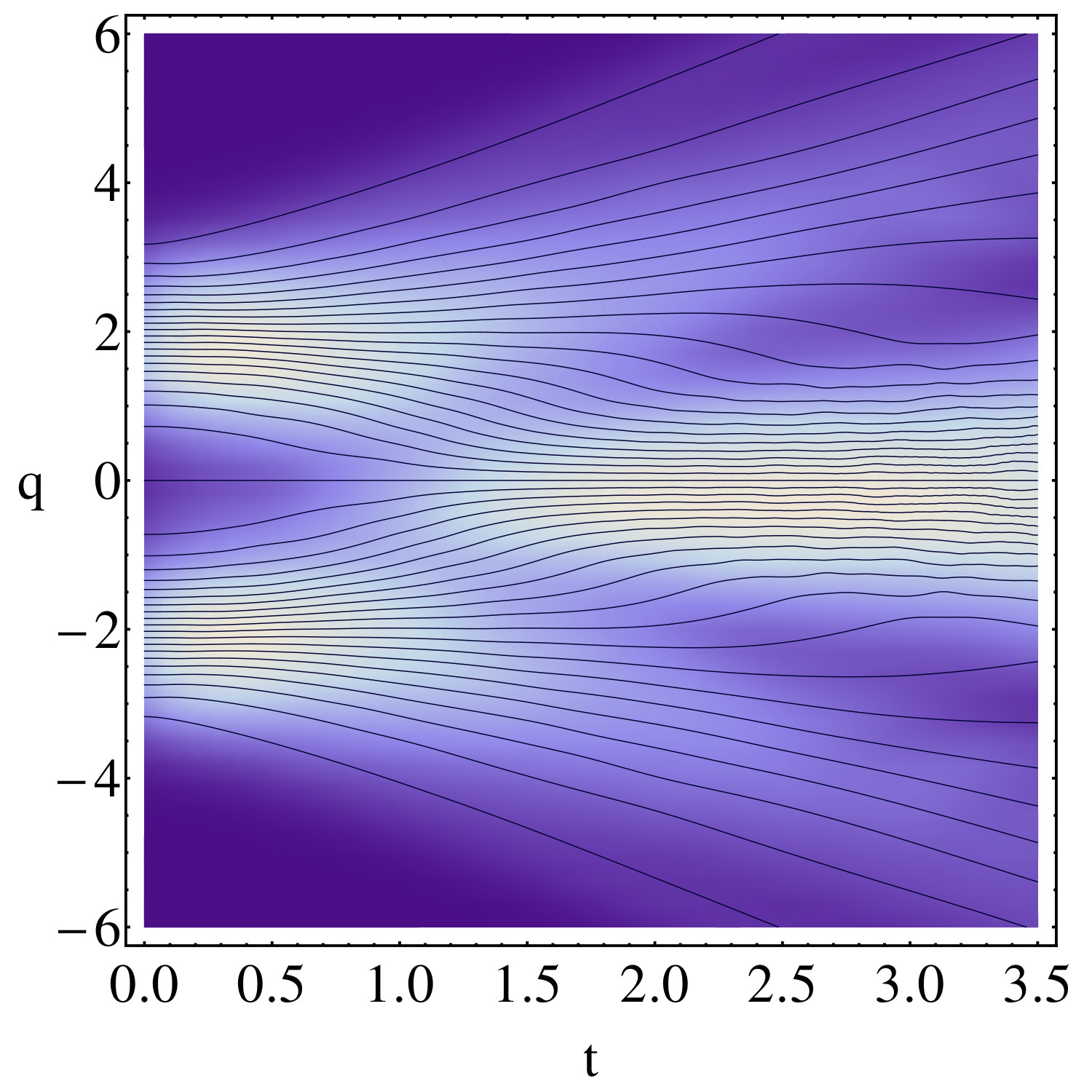}
        \includegraphics[width=0.47\textwidth]{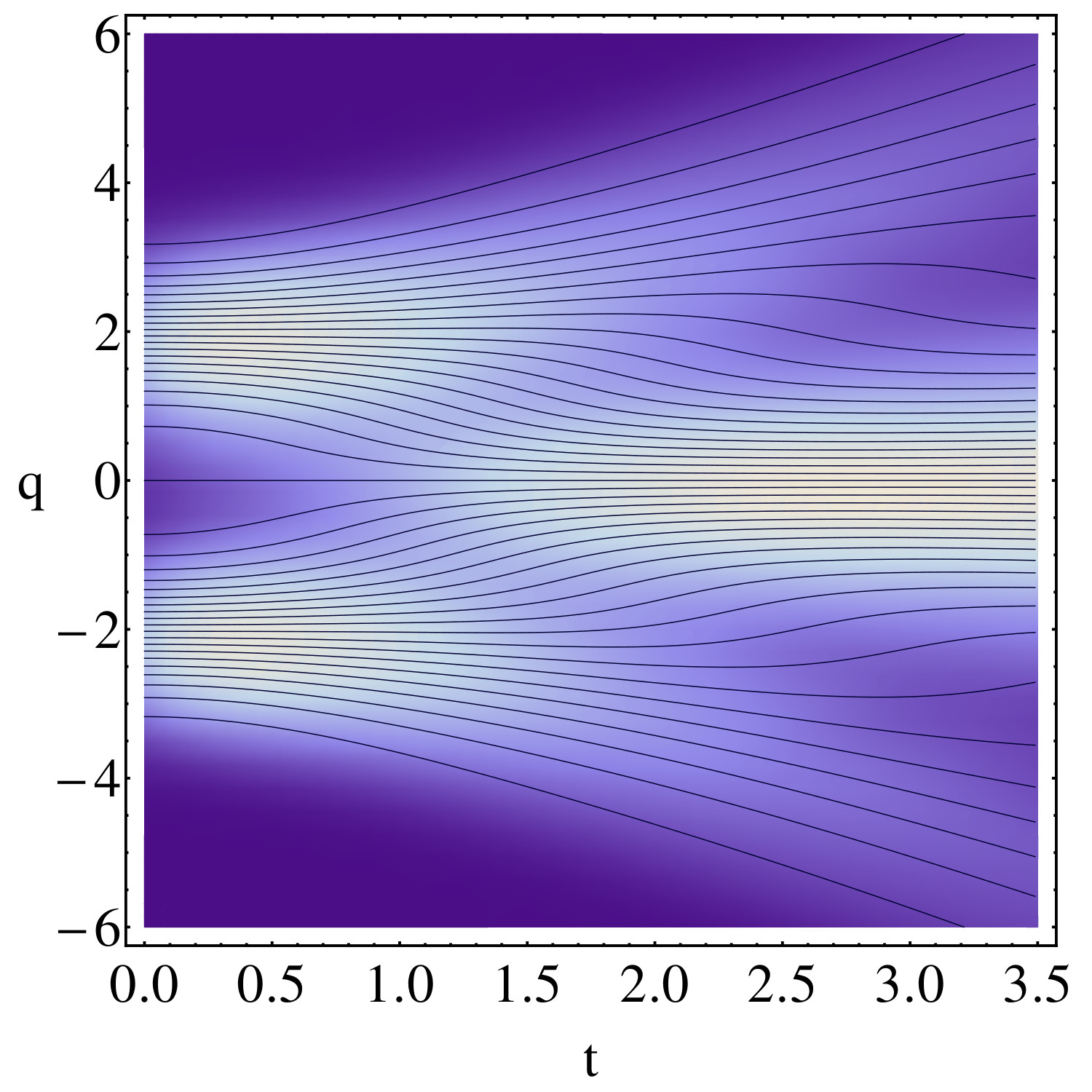}
    \end{center}
    \caption{\label{dynamics} (Color online).
       Trajectory and (smoothed) density plots
        of the computed MIW (top) and dBB (bottom) world 
        trajectories 
    for two identical initial Gaussian wavepackets in units of the initial
spread (\ie~$\sigma=1$) and $N=41$ worlds; time is given in units of
$\hbar/(2m)$.\blk}
\end{figure}

\begin{figure}[!t]
	\includegraphics[width=0.47\textwidth]{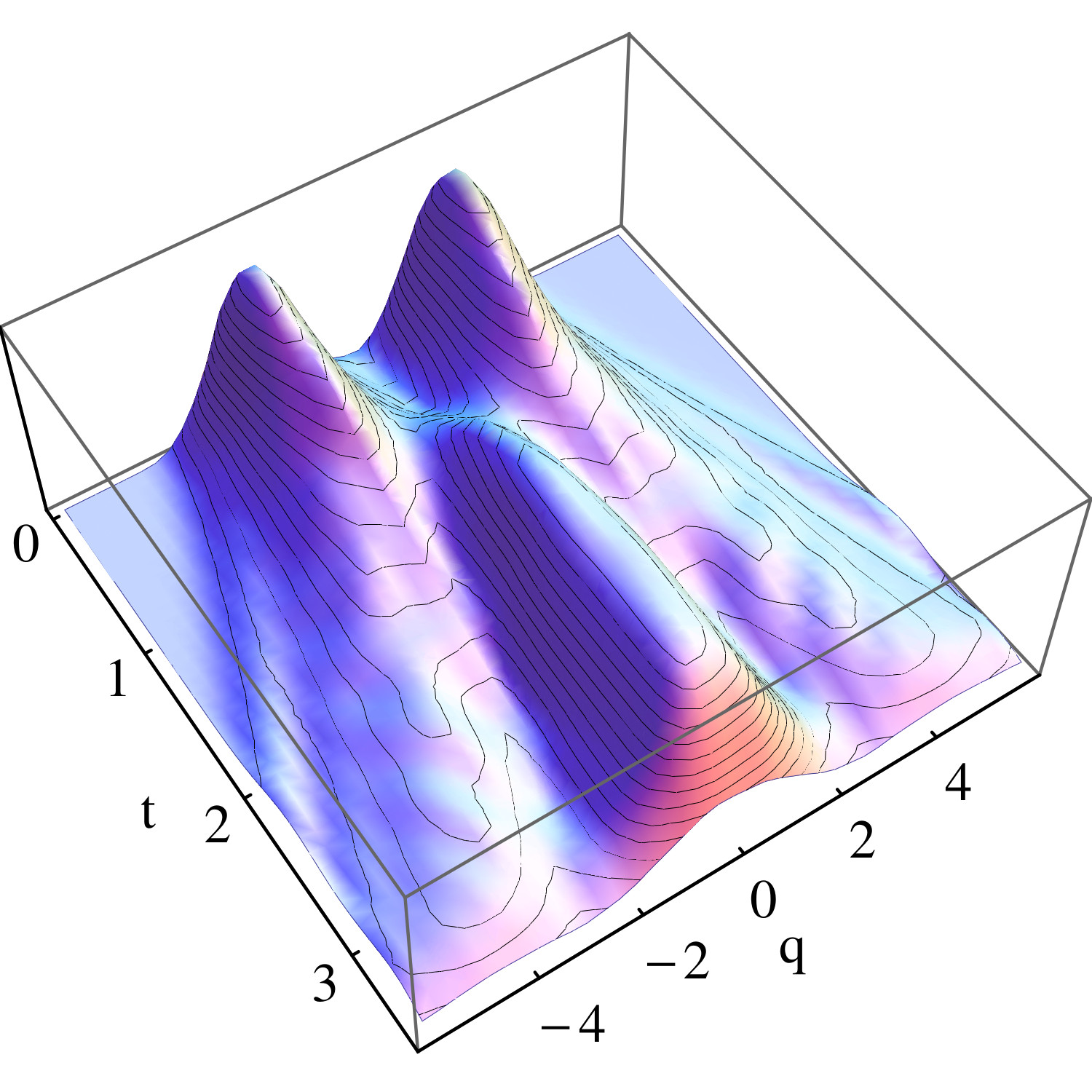}
	\includegraphics[width=0.47\textwidth]{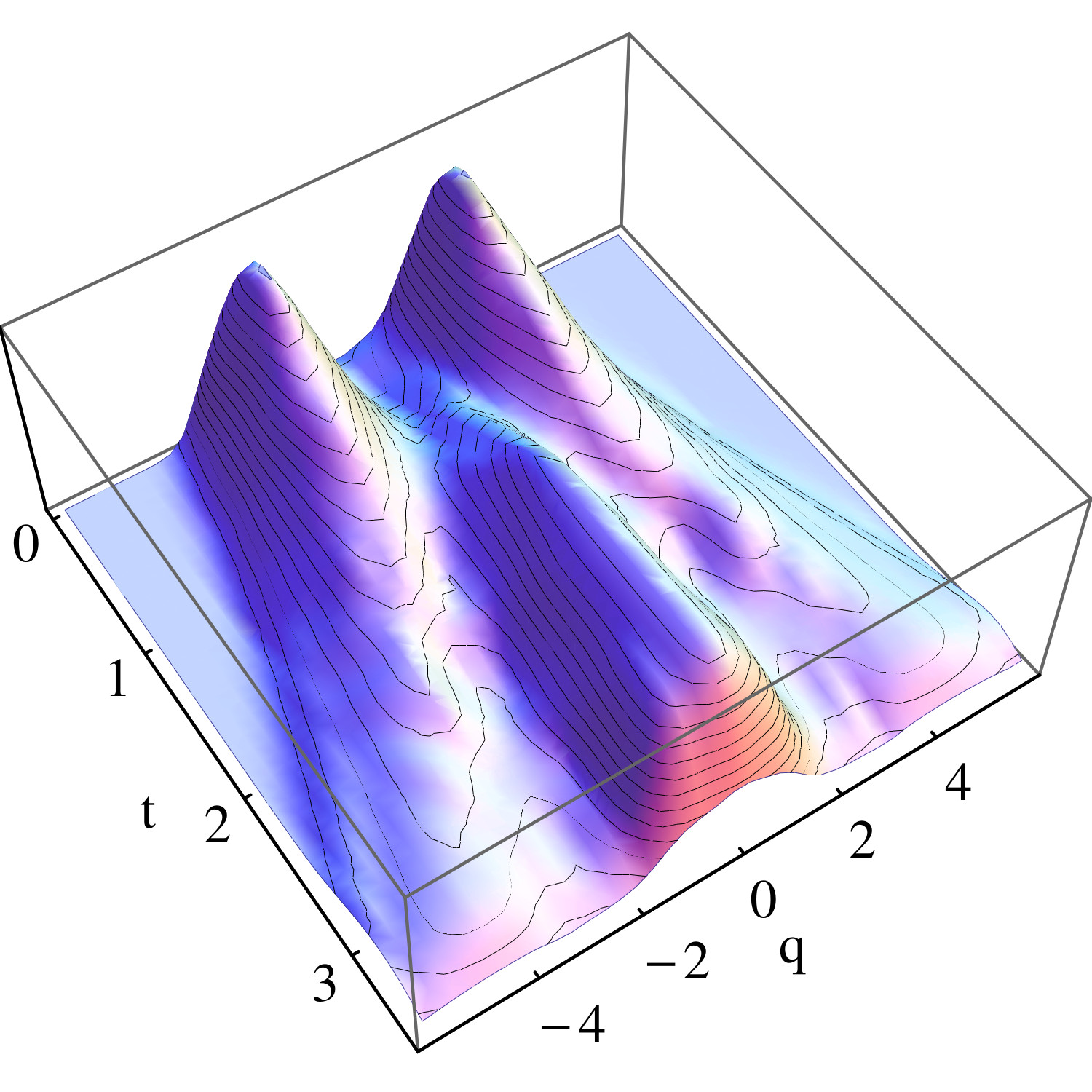}
    \caption{\label{dynamics3d} (Color online).
     Smoothed histogram
        plots of the computed MIW (top) and dBB (bottom) world trajectories for
        two identical initial Gaussian wavepackets in units of the initial spread (\ie~$\sigma=1$) \blk  and $N=41$
    worlds; time is given in units of $\hbar/(2m)$.\blk}
\end{figure}

In its simplest form, the `double-slit' scenario comprises the free 
evolution of a 1-dimensional \blk wave function $\Psi_t$ for an initial value $\Psi_0$ given by a
symmetric  superposition of two identical separated wavepackets.  
We chose  \blk real, Gaussian wavepackets with spread $\sigma=1$
and initial separation of $4$ units; see Fig~\ref{qm}. 
As described in Sec.~\ref{sec:limit}, the corresponding initial values for
the MIW approach are $N$ worlds,
${X}(0)=({x}_1(0),\ldots,{x}_N(0))$, distributed according
to $|\Psi_0|^2$ with zero initial velocities.
In order to numerically integrate the MIW equations of motion Eq.~\ref{eq:MIW}
to find the corresponding world-particle trajectories ${X}(t)$ we employed
 a standard Verlet integration scheme  \cite{verlet}. 
In our example we used $N=41$. The computed trajectories ${X(t)}$
are shown in the top two plots of Fig.~\ref{dynamics}  and
Fig.~\ref{dynamics3d}  using a linearly smoothed
density and histogram plot.  For the comparison with the exact quantum solution $\Psi_t$
we computed the corresponding Bohmian trajectory $x^{\mathrm{dBB}}_i(t)$
according to Eq.~\ref{velocityS} for each initial world configuration $x_i$ for
$i=1,2,\ldots,N$.  These trajectories are shown in two bottom plots again using
a linearly smoothed density and histogram plot.  
Note that due to equivariance, the Bohmian
configurations ${\bf x}^{\mathrm{dBB}}_{i}(t)$ are distributed according to
$|\Psi_t|^2$ for all times $t$ as the ${\bf x}^{\mathrm{dBB}}_i(0)$ are distributed according to
$|\Psi_0|^2$. Therefore, the bottom  density plot  in Fig.~\ref{dynamics} is a discretized plot of
$|\Psi_t|^2$ as shown in Fig.~\ref{qm}  (this discretization, corresponding
to using just $N=41$ Bohmian worlds and linear smoothing, is responsible for the
blurriness relative to Fig.~\ref{qm}). 

By comparison of the plots we conclude that the MIW approach is at least
qualitatively able \blk to reproduce quantum interference phenomena. Furthermore, it should be
stressed that with respect to usual numerical PDE techniques on grids the MIW
approach
could have several substantial computational advantages: As in dBB,
the worlds ${X}(t)$ should remain approximately $|\Psi_t|^2$-distributed \blk  for all
times $t$ which implies that without increasing $N$, the density is naturally
well-sampled in regions of high density even when the support of
$\Psi_t$ becomes very large. The computational effort is then automatically
focussed on
high-density regions of configuration space, \ie, on the crucial regions where numerical
errors have to be minimized to ensure convergence in the physically relevant $L^2$ norm.
Furthermore, transmission and reflection coefficient can be computed simply by counting
worlds. These advantageous also hold true for methods of integration the \sch\ 
equation on co-moving grids as, e.g., proposed by Wyatt et al. \cite{wyatt}. However,
there are reasons to expect \blk that the MIW approach may be \blk
 numerically more stable (see Ref.~\cite{deckert} for a discussion of such numerical instabilities 
 in conventional approaches) \blk and less performance intensive.

 There is much work to be done on the question of time evolution using our 
 MIW approach. Even in the example above, the  convergence 
 as a function of $N$, the number of worlds, remains to be investigated. 
 Also, there is the issue of wave functions 
 with nodes (or even finite regions of zero values). These allow for 
 different wave functions $\Psi_0(q)$ having identical $P_0(q)$, and  identical  $\nabla  S_0(q)$ everywhere that  $P_0(q)\neq 0$.  From  Sec.~\ref{sec:limit}
 these different wave functions will   all generate  the same initial conditions for our 
 ensemble of many worlds,  for any finite value of $N$,   and thus are indistinguishable in the MIW approach,  
 contrary to quantum predictions (see also Sec.~10 of \cite{sebens}). \blu For example,  the wavefunctions $\Phi_0(q)$ and $|\Phi_0(q)|$ are of this type, where  $\Phi_0(q)$ is any real wavefunction that takes both positive and negative values. 
\blk
 However it is plausible that restricting the  moments of the  quantum energy to be finite would  typically  eliminate
 all, or all but one, of such wavefunctions,  by enforcing the continuity of $\nabla \Psi_0$ \blu everywhere \blk \cite{energy1,energy2}.    Finally,  beyond the toy 
 model, there is the question of whether the general approach of many interacting worlds 
allows  successful numerical treatments of  more  general
situations, with common  potentials and $K=DJ>1$.

\section{Conclusions}

We have introduced an approach to quantum phenomena in which all quantum effects
are due to interactions between a large but finite number $N$ \blk of worlds, and
probabilities arise from assigning an equal weighting to each world.  A number
of generic quantum effects, including  wavepacket  spreading,  
tunnelling, zero
point energies, and interference \blk have been shown to be consequences of the mutual repulsion
between worlds   \blk (Sec.~IV).  This
alternative realistic interpretation of quantum phenomena is also of interest in
not requiring the concept of a wave function. 
The \blk wave function does not
exist other than as an epiphenomenon  in the notional limit 
that the initial distribution of worlds approaches $|\Psi_0(x)|^2$ 
{\it and} the initial velocity of each world approaches Eq.~(\ref{velocityS}) as $N\to\infty$.

 For finite $N$, our MIW approach can only ever give an approximation to 
quantum mechanics, but since quantum mechanics
is such an accurate theory for our observations, we require this approximation to 
be very good. \red In this context, it is worth   revisiting the question 
raised in Sec.~\ref{sec:limit}: \blk 
what restrictions must one place on the initial distribution of world configurations 
and velocities so that nearly quantum behaviour (that is, observations consistent with 
quantum mechanics for {\em some} wavefunction) 
will  be experienced by macroscopic observers in almost all worlds?    
The answer may depend on the details of the inter-world potential, 
\red but we can suggest directions to explore.

\red As discussed in Sec.~\ref{sec:limit}, it \blk may be necessary 
to impose the quantum velocity condition of Eqs.~(\ref{velocityS}) and (\ref{sdef}) \blk using {\em some} smoothly 
varying (on the scale of the separation between nearby worlds) 
single-valued complex function \red $\Psi_\tau$ \blk on configuration space.
\red However, \blk it is conceivable that \ other (perhaps even most, by some measure 
of typicality)  initial conditions would relax, under our 
interacting-world dynamics, to conditions approximating quantum behaviour, 
at least at the scale which can be probed by a macroscopic observer. 
This is one of the big questions which remains to be investigated. \red One might 
think that this idea would never work, because the velocities 
would relax to \blu some type of random Maxwellian distribution. \red 
 However, this may well not be the case, because \blu 
 the many-body interaction potentials and forces in our MIW approach 
--- which generically reproduce various quantum phenomena as per Secs.~IV-VI \red --- are very 
different from those assumed in classical statistical mechanics \blu (see also the `gas' example discussed in Sec.~I~A).   \blk

Other matters for future investigation include   how spin and   entanglement
\blu phenomena such as teleportation and Bell-inequality violation \blk are
modelled in the
MIW approach. The latter \blk will require studying the case of worlds with configuration
spaces of at least two dimensions (corresponding to two one-dimensional
systems), and will also allow analysis of the quantum measurement problem (where
one system acts as an `apparatus' for the other). This may help clarify 
the ontology and epistemology of any fundamental new theory based upon
the MIW approach to quantum mechanics. \blk 

In the context of entanglement, it is worth comparing our MIW approach with 
conventional  many-worlds approaches. The latter  are often motivated by the desire,
first,  to restrict reality to only the wave function, and second, to avoid the 
explicit nonlocality which arises from entanglement \blk in other realist versions of quantum mechanics. Our
approach is, by  contrast,   motivated by the desire to 
eliminate the wave function.  It  furthermore
elevates the  nonlocality of quantum mechanics to a kind of
``super-nonlocality'':  particles in different worlds are nonlocally connected through the proposed MIW interaction, thus leading, indirectly, to  nonlocal interactions between particles in the same world. 
\blk

 Turning from questions of foundations and interpetations to applied science, 
the MIW approach  provides  a promising controlled
approximation for simulating \blk quantum groundstates and the time-dependent  Schr\"odinger equation, as
discussed in Secs.~\ref{sec:groundstates} and \ref{sec:quantum-evolution}. 
In particular, we have shown that it is able to reproduce quantum interference phenomena, 
at least qualitatively. Quantitative comparisons with different initial conditions,  convergence as a function of
 the number of worlds  $N$, and generalizations to higher dimensions, is a matter for 
immediate  future work.

\acknowledgments{We would like to thank  S. Goldstein and  C.
    Sebens for valuable discussions,
    and A. Spinoulas for independently verifying the calculations in Sec.
    VI. 
   Furthermore, D.-A.D.\ thanks Griffith University for its
   hospitality and gratefully acknowledges funding by the AMS-Simons Travel Grant. \blk

\appendix

\section{Connection of Bohmian force with the Hamiltonian formulation of
MIW}\label{apx:A}

We briefly note here how the interworld potential $U_N$ in Sec.~II~C is related to a corresponding approximation of the quantum potential $Q$ in Sec.~II~A.  We also exhibit a direct correspondence between the interworld force ${\bf r}_N({\bf q};{\bf X}_t)$ and the Bohmian force ${\bf r}_t({\bf q})$ for the case of the  toy 1D model defined in Sec.~III.

For any particular wave \blk function \blk $\Psi_t({\bf q})$, the dBB equation of motion (\ref{Bohmacc}) for the world-particle is generated by the Hamiltonian
\[ H^{\Psi_t}({\bf x},{\bf p}) := \sum_{k=1}^K \frac{(p^k)^2}{2m^k} + V({\bf x}) + Q_t({\bf x}), \]
where $Q({\bf q})$ is the quantum potential corresponding to $\Psi_t({\bf q})$ [identifying the momentum component $p^k$ with the right hand side of Eq.~(\ref{velocityS})].  Hence, the evolution of $N$ Bohmian world-particles, each guided by $\Psi_t({\bf q})$, are precisely described via the time-dependent Hamiltonian
\begin{align*} 
H_N^{\Psi_t}({\bf X},{\bf P}) &:= \sum_{n=1}^N H^{\Psi_t}({\bf x}_n,{\bf p}_n)\\ 
&=  \sum_{n=1}^N \sum_{k=1}^K \frac{(p^k_n)^2}{2m^k}+ \sum_{n=1}^N V({\bf x}_n) + \sum_{n=1}^N Q_t({\bf x}_n) .
\end{align*}
Thus, if $Q_t({\bf q})$ (and its gradient) can be approximated by some function $\tilde Q({\bf q}; {\bf X}_t)$, depending on the configurations ${\bf X}_t$ of the $N$ worlds (assumed to sample $|\Psi_t({\bf q})|^2$), then one immediately has a suitable corresponding Hamiltonian formulation of the MIW approach, with an interworld potential $U_N$ in Eq.~(\ref{ham}) of the alternative form
\begin{equation} \label{altern}
U_N({\bf X})=\sum_{n=1}^N \tilde Q({\bf x}_n; {\bf X}) 
\end{equation} 
to that in Eq.~(\ref{un}).

To show how the above form is related to the positive definite form in Eq.~(\ref{un}), consider the relation
\[ \int d{\bf x} \,|\Psi({\bf q})|^2 Q({\bf q}) = \int d{\bf q}\, |\Psi({\bf q})|^2 \sum_{k=1}^K\frac{\hbar^2}{8m^kP^2}\left(\frac{\partial P}{\partial q^k}\right)^2 \]
following from integration by parts and Eq.~(\ref{eq:quantum-potential}) \cite{dbb}.
In particular, since the $N$ world-particles have configurations sampled according to $|\Psi({\bf q})|^2$, it immediately follows that
\begin{align*} 
\frac{1}{N} \sum_{n=1}^N Q({\bf x}_n) &\approx \frac{1}{N} \sum_{n=1}^N \sum_{k=1}^K\frac{\hbar^2}{8m^kP^2}\left.\left(\frac{\partial P}{\partial q^k}\right)^2\right|_{\blk q_k = x_n^k}.
\end{align*}
Hence, $U_N$ in Eq.~(\ref{altern}) may alternatively be replaced by 
the form in  Eq.~(\ref{un}), corresponding to a suitable approximation of $P$ and its derivatives in terms of the configurations of the $N$ worlds.  \blk Here ``suitable'' means that we must check that the \blk corresponding MIW force is given by ${\bf r}_N({\bf x}_n; {\bf X})=-\nabla_{{\bf x}_n}\, U_N({\bf X})$ \blk in the limit $N\to\infty$. \blk 

For the \blk toy MIW model \blk of Sec.~III,  we can \blk demonstrate  directly that the force in  Eq.~(\ref{hamforce}) is an approximation of the Bohmian force, as per Eq.~(\ref{eq:quantum-force}) of Sec.~II. To so so, we first note that, for the one-dimensional case, the derivative of any regular function ${\blk \varphi}(x)$ can be approximated at $x=x_n$ by either of
\[{\blk \varphi}'(x_n)\approx \frac{{\blk \varphi}(x_{n+1})-{\blk \varphi}(x_n)}{x_{n+1} - x_{n}}\approx \frac{{\blk \varphi}(x_{n})-{\blk \varphi}(x_{n-1})}{x_{n} - x_{n-1}} ,\]
for sufficiently large $N$.
Hence, using Eq.~(\ref{papprox}), one obtains
\begin{equation} \label{fderiv}
\left[\frac{1}{P(x_n)}\frac{d}{dx} \right]{\blk \varphi}(x_n) \approx ND_+ {\blk \varphi}(x_n) \approx N D_- {\blk \varphi}(x_n),
\end{equation}
where the difference operators $D_\pm$ are defined by 
\[D_+{\blk \varphi}_n:={\blk \varphi}_{n+1}-{\blk \varphi}_n,~~~~~D_-{\blk \varphi}_n:= {\blk \varphi}_n-{\blk \varphi}_{n-1}\]
for any sequence $\{{\blk \varphi}_n\}$.  

Now, as may be checked by explicit calculation from the definition of the quantum potential in Eq.~(\ref{eq:quantum-potential}), for a one-dimensional particle the Bohmian force in Eq.~(\ref{Bohmforce}) can be written in the form 
\begin{equation} \label{niceforce}
r_t=(\hbar^2/4m)(1/{P})\left[ P (P'/{P})'\right]',
\end{equation} 
\ie,
\begin{equation} \nn
r_t(q) = \frac{\hbar^2}{4m} \left[\frac{1}{P(q)}\frac{d}{dq}\right] P(q)^2 \left[\frac{1}{P(q)}\frac{d}{dq}\right]^2 P(q) ,
\end{equation}
where the derivative operators act on all terms to their right.  Applying Eq.~(\ref{fderiv})
 then gives
\begin{align} \nn
r_t(x_n) &\approx \frac{\hbar^2}{4m} N^3\,D_+\left[ P(x_n)^2 D_+ D_- P(x_n)\right]\\ \nn
&\approx \frac{\hbar^2}{4m} D_+\left[ \frac{1}{(x_n-x_{n-1})^2} D_+ D_- \frac{1}{x_n-x_{n-1}}\right]\\ 
&= r_N(x_n; {\blk X})
\end{align}
as desired, where the second line uses Eq.~(\ref{papprox}), and the last line follows by expansion and direct comparison with Eq.~(\ref{hamforce}).

\section{Lower bound for the interworld potential}\label{apx:D}

To obtain the bound in Eq.~(\ref{enboundn}), first let $f_1,\dots, f_N$ and $g_0,g_1,\dots,g_N$ be two sequences of real numbers such that $g_0=g_N=0$.  
It follows that
\[ \sum_{n=1}^N f_n(g_n-g_{n-1})= -\sum_{n=1}^{N-1} (f_{n+1}-f_n)g_n, \]
as may be checked by explicit expansion.
Further, the Schwarz inequality gives
\[ \left[ \sum_{n=1}^N f_n(g_n-g_{n-1})\right]^2 \leq \left[\sum_{n=1}^N (f_n)^2\right]\,\left[\sum_{n=1}^N (g_n-g_{n-1})^2\right] ,\]
with equality if and only if $f_n=\alpha (g_n-g_{n-1})$ for some $\alpha$.  Combining these results then yields
\[ \sum_{n=1}^N (g_n-g_{n-1})^2 \geq \frac{\left[\sum_{n=1}^{N-1} (f_{n+1}-f_n)g_n\right]^2}{\sum_{n=1}^N (f_n)^2} . \]
The particular choices $f_n=x_n-\langle x\rangle_N$ and $g_n=1/(x_{n+1}-x_n)$ (with $x_0=-\infty$ and $x_{N+1}=\infty$) immediately yield, using the definition of $U_N$ in Eq.~(\ref{un1d}), the lower bound
\beq \label{unbound}
U_N \geq \frac{\hbar^2}{8m}  \frac{\left[\sum_{n=1}^{N-1} 1\right]^2}{\sum_{n=1}^N (x_n-\langle x\rangle_N)^2} = \frac{(N-1)^2}{N}\frac{\hbar^2}{8m}\frac{1}{\blk {\cal V}_N}. \eeq
for the interworld potential energy, with equality if and only if the Schwarz inequality saturation condition
\begin{equation} \label{sat}
x_n -\langle x\rangle_N = \frac{\alpha}{x_{n+1} - x_n} - \frac{\alpha}{x_{n} - x_{n-1}} 
\end{equation}
is satisfied.  Eq.~(\ref{enboundn}) \blk then follows via Eq.~(\ref{ham}), as desired.  

Similar bounds can be obtained for other choices of the interworld potential, but will not be discussed here.

\section{Groundstate energy and configuration of toy model
oscillator}\label{apx:groundstates}

To derive the groundstate energy in Eq.~(\ref{enosc}), for the 1D toy model oscillator, note first that for a harmonic oscillator potential $V(x)=(1/2)m\omega^2 \blk x^2$  one has
\bqa \nn
\langle E\rangle_N &=& N^{-1}H_N \geq  N^{-1} \sum_{n=1}^N V(x_n) + N^{-1} U_N(x)\\
\nn &\geq&  \frac{1}{2} m\omega^2\, {\blk {\cal V}_N} + \left(\frac{N-1}{N}\right)^2 \frac{\hbar^2}{8m}\frac{1}{\blk {\cal V}_N}\\
\nn &=& \frac{N-1}{N} \frac{\hbar\omega}{4} \left[ \frac{2Nm\omega\,{\blk {\cal V}_N}}{(N-1)\hbar}  +   \frac{(N-1)\hbar}{2Nm\omega\,{\blk {\cal V}_N}} \right]\\
\nn &\geq& \frac{N-1}{N} \frac{\hbar\omega}{2},
\eqa
The second line follows via $\langle \blk x^2 \blk\rangle_N\geq \blk {\cal V}_N$ and inequality (\ref{unbound}), and the last line via $z+1/z\geq 2$. 

To show this inequality chain can be saturated, as per Eq.~(\ref{enosc}), the conditions for equality must be checked at each step.  It follows that the lower bound is achievable if and only if (i) the momentum of each world vanishes, \ie, $p_1=\dots=p_N=0$; and (ii) the configuration satisfies both
\beq \label{cond}
\langle x \rangle_N=0,~~~ {\blk {\cal V}_N} = \frac{N-1}{N}\, \frac{\hbar}{2m\omega} ; \eeq
and the Schwarz inequality saturation condition \label{schwarz}
\begin{equation} \label{recur}
x_n = \frac{\alpha}{x_{n+1} - x_n} - \frac{\alpha}{x_{n} - x_{n-1}} 
\end{equation}
from Eq.~(\ref{sat}), for some constant $\alpha$ (using $\langle x
\rangle_N=0$).  These conditions generate a 2nd-order recurrence relation with
fixed boundary conditions, yielding a unique solution for any given number of
worlds $N$. The corresponding unique groundstate configurations converge to the
quantum Gaussian groundstate wave function for $N\to\infty$, as discussed in
\grn Secs.~II~C and \blk V~A.

To determine the groundstate configuration, for a given number of worlds $N$, note that summing each side of Eq.~(\ref{recur}) from $n=1$ up to any $n<N$ gives
\begin{equation} \label{recur2}
\frac{1}{x_{n+1}-x_n} = \frac{x_1+\dots+x_n}{\alpha} . 
\end{equation}
This is also satisfied for $n=N$ (defining  $x_{N+1}=\infty$ as usual), since $x_1+\dots+x_N=0$ from Eq.~(\ref{cond}).  Further, noting that $x_1<\dots<x_N$ and $(x_1+\dots+x_n)/n\leq \langle x\rangle=0$, it follows from Eq.~(\ref{recur2}) that $\alpha<0$.  Hence, since summing the squares of each side of Eq.~(\ref{recur}) over $n$ gives
\[ N\, {\blk {\cal V}_N} = \frac{8m\alpha^2}{\hbar^2} U_N , \]
and inequality (\ref{unbound}) is saturated by the groundstate, it follows via Eq.~(\ref{cond}) that 
\begin{equation} 
\alpha = -\frac{N\, {\cal V}_N}{N-1} = -\frac{\hbar}{2m\omega}.
\end{equation}

Defining the dimensionless configuration coordinates
\begin{equation} \label{qndefD}
{\blk \xi}_n := (-\alpha)^{-1/2} \,x_n = \sqrt{2m\omega/\hbar} \,\,x_n , 
\end{equation}
conditions (\ref{cond}), (\ref{recur}) and (\ref{recur2}) then simplify to
\begin{equation} \label{cond2D}
{\blk \xi}_1+\dots+{\blk \xi}_N = 0,~~~({\blk \xi}_1)^2+\dots ({\blk \xi}_N)^2 = N-1,
\end{equation}
\begin{equation}
\label{recur3D}
{\blk \xi}_{n+1} = {\blk \xi}_n - \frac{1}{{\blk \xi}_1+\dots+{\blk \xi}_n}.
\end{equation}
Since these equations are invariant under ${\blk \xi}_n\rightarrow-{\blk \xi}_n$, the symmetry property 
\begin{equation} 
{\blk \xi}_{n}=-{\blk \xi}_{N+1-n}  
\end{equation}
follows from the uniqueness of the groundstate.

\blk For $N \in \{2,3,4,5\}$, recurrence relation (\ref{recur3D})  
reduces to solving an equation no more than quadratic in \blu $({ \xi}_1)^2$, \blk allowing the groundstate 
to be obtained analytically. For example, for $N=3$ one finds 
\[{ \xi}_1=-1, ~~~{ \xi}_2=0, ~~~ { \xi}_3=1, \] 
while for $N=4$ one obtains 
$$ { \xi}_1 = -\frac{\sqrt{7+\sqrt{17}}}{2\sqrt{2}}=-{ \xi}_4,~ { \xi}_2={ \xi}_1+\frac{1}{2}\sqrt{7-\sqrt{17}}=-{ \xi}_3 . $$ \\
 \blk More generally, it may be used to express ${\blk \xi}_2,\dots,{\blk \xi}_{[N/2]}$ in terms of ${\blk \xi}_1$, where $[z]$ denotes the integer part of $z$.  The groundstate configuration may then be numerically determined by solving the condition $({\blk \xi}_1)^2+\dots+({\blk \xi}_{[N/2]})^2=(N-1)/2$ for ${\blk \xi}_1$ (following from Eq.~(\ref{cond2D}) and the above symmetry property).

\end{document}